\documentclass[11pt]{article}
\usepackage{jair,theapa,rawfonts}
\usepackage[utf8]{inputenc}
\usepackage{amsfonts}
\usepackage{amsmath}
\usepackage{amssymb}
\usepackage{array}
\usepackage{amsthm}
\usepackage{subfigure}
\usepackage{mathdots}
\usepackage{algorithm, algorithmic}
\usepackage[table,x11names]{xcolor}
\usepackage{color}
\usepackage{pgf}

\usepackage{hyperref}
\usepackage{makecell}
\usepackage{authblk}

\usepackage[round]{natbib}   
\newcommand{\R}{\mathbb{R}}
\newcommand{\PSNR}{\text{PSNR}}
\newcommand{\MSE}{\text{MSE}}
\newcommand{\SSIM}{\text{SSIM}}
\newcommand{\MAX}{\text{MAX}}

\title{A study of why we need to reassess full reference image quality assessment with medical images}

\author[1,2]{\small  Anna Breger}
\author[1]{Ander Biguri}
\author[3]{Malena Sabaté Landman}
\author[4]{Ian Selby}
\author[5]{Nicole Amberg}
\author[2]{Elisabeth Brunner}
\author[6,7]{Janek Gröhl}
\author[8,9]{Sepideh Hatamikia}
\author[2]{Clemens Karner}
\author[10]{Lipeng Ning}
\author[1]{Sören Dittmer}
\author[1]{Michael Roberts}
\author[11]{AIX-COVNET Collaboration}
\author[1]{Carola-Bibiane Schönlieb\footnotesize}

\affil[1]{University of Cambridge, Department of Applied Mathematics and Theoretical Physics, Cambridge, UK}
\affil[2]{Medical University of Vienna, Center of Medical Physics and Biomedical Engineering, Vienna, Austria}
\affil[3]{Emory University, Department of Mathematics, Atlanta, United States}
\affil[4]{University of Cambridge, Department of Radiology, Cambridge, United Kingdom}
\affil[5]{Medical University of Vienna, Department of Neurology, Vienna, Austria}
\affil[6]{University of Cambridge, Department of Physics, Cambridge, United Kingdom}
\affil[7]{Cancer Research UK, Cambridge Institute, University of Cambridge, United Kingdom}
\affil[8]{Research Center for Clinical AI-Research in Omics and Medical Data Science (CAROM), Department of Medicine, Krems an der Donau, Austria}
\affil[9]{Austrian Center for Medical Innovation and Technology, Wiener Neustadt, Austria}
\affil[10]{Harvard Medical School, Brigham and Women's Hospital, Boston, MA, United States}
\affil[11]{A list of members is provided in the acknowledgments.}

\begin{document}

\maketitle

\begin{abstract}
Image quality assessment (IQA) is indispensable in clinical practice to ensure high standards, as well as in the development stage of machine learning algorithms that operate on medical images. The popular full reference (FR) IQA measures PSNR and SSIM are known and tested for working successfully in many natural imaging tasks, but discrepancies in medical scenarios have been reported in the literature, highlighting the gap between development and actual clinical application. Such inconsistencies are not surprising, as medical images have very different properties than natural images, and PSNR and SSIM have neither been targeted nor properly tested for medical images. This may cause unforeseen problems in clinical applications due to wrong judgment of novel methods. This paper provides a structured and comprehensive overview of examples where PSNR and SSIM prove to be unsuitable for the assessment of novel algorithms using different kinds of medical images, including real-world MRI, CT, OCT, X-Ray, digital pathology and photoacoustic imaging data. Therefore, improvement is urgently needed in particular in this era of AI to increase reliability and explainability in machine learning for medical imaging and beyond. Lastly, we will provide ideas for future research as well as suggesting guidelines for the usage of FR-IQA measures applied to medical images. 
\end{abstract}

\section{Introduction}
Advances in medical imaging technologies have been groundbreaking in the last decades, ranging from new modalities of scanners, including hardware innovations, and advances in mathematical tools for image reconstruction to the current state of the art in machine learning techniques. The overall aim is to apply novel technology in clinical scenarios to improve patient's care. In order to ensure a clinically acceptable quality of the novel imaging techniques, quantitative image quality assessment (IQA) plays an important role for quality assurance in addition to visual inspections. 

Quantitative IQA can roughly be divided into three categories based on their underlying assumptions and the available information for their evaluation, cf.~\citep*{iqasurv}. The first one is called full reference (FR) IQA, where a full known image is used as a reference (or ground truth) and the quality of a given image is evaluated in a comparative way that relies on a meaningful notion of distance between the two images.
No reference (NR) IQA, on the other hand, is not based on a one-to-one image comparison, and instead it aims to judge the quality of a given image solely by evaluating properties.  
Lastly, reduced reference (RR) IQA lays somewhere in between, for example, using specific retrieved image information such as edge information or local image characteristics as a reference. Most commonly, NR- and FR-IQA measures have been developed and used to solve quite different problems. As FR-IQA requires a reference image, it can only be used in very specific tasks. This includes the evaluation of novel (traditional or machine learning-based) imaging methods in their development stage and experiment calibration, where reference data is available. In this case, the measure is used to make conclusions about the performance of the algorithms or settings in different imaging tasks, such as image compression, denoising or reconstruction, and, consequently, to reinforce the success of the different methods. NR-IQA, on the other hand, is used to extract and judge quality information from a given image on the spot, particularly when there is no access to the reference image. 
This is usually the case when evaluating the quality of a given image outside of the development stage, such as real-time quality control of image acquisition in a hospital.  

In this paper, we focus on FR-IQA measures that have been broadly used for the evaluation of novel image processing algorithms that are operating on medical images. 
Research on the development of novel methodologies, mainly driven within the areas of mathematics, computer science and engineering, often uses FR-IQA for the evaluation of the proposed algorithms. The performance of the methods under these metrics influences which methods are published, therefore, the choice of FR-IQA at the developing stage has a capital influence on the method choices that are subsequently available.  
However, the authors carrying out research at the development stage are not necessarily experts on particular applications and, therefore, might not take into account the specific nuisances of medical imaging data and the importance of corresponding IQA measure choice. For this reason, the field might be promoting new methodologies which are not the most suited for clinical tasks, such as diagnosis of specific data. 

Some fundamental questions come alongside the task of evaluating novel methodologies and their application. Could or should the development of novel methods be decoupled from their application potential? Is it feasible to develop highly sophisticated algorithms and assess their applicability without an expert's opinion? And, on the other hand, is it feasible to assume that an application expert needs to be available every time a new methodology is developed?
In the last decades, these unsolved principles have led to rather disconnected research areas of model development and eventual application. This is particularly true in the fast advancing domain of machine learning, where publicly shared data is presumably enabling the development of novel algorithms for specific applications more easily, but where the lack of direct contact with experts on the data is hindering their real applicability, e.g.~in medical imaging. In that case, accurate evaluation with IQA measures becomes even more important as visual inspection from experts is not feasible. 

However, the most commonly employed FR-IQA metrics have not been developed nor broadly tested to work successfully for medical images. Usually, novel quality measures are tested by computing correlations of their outcome to the mean opinion scores of experts who have manually rated the images. In order to do so, there are several standard data bases that provide such rated data, see e.g.~the LIVE data base \citep{live} or the image database TID2012 \citep{PONOMARENKO201557}. However, these kinds of annotations are rare, as they are very time consuming and task-dependent. For medical images, there is another layer of added difficulty: it is very hard to publish clinical images due to the sensitivity of the data. For these reasons, currently and to the best of our knowledge, there has been no publicly available database with FR ratings for medical images to assess the performance of the quality measures. Thereto, we have recently published a data set with photoacoustic images and expert annotations, which are available on Zenodo \cite{breger_2024_13325197}. 
Previously published studies including medical in-house data have raised concerning results \cite{10040654, breger2024study2}. The data sets include ratings by experts and radiologists, showing that the most common FR-IQA measures perform poorly for the studied tasks. 

With that, this paper has the following main objectives:
\begin{enumerate}
    \item Providing a structured collection of pitfalls of standard FR-IQA measures (namely PSNR and SSIM, as well as the more recently introduced LPIPS) when used for common medical imaging tasks. We show real-world examples in different medical imaging applications. The examples are described in detail, because without in-depth discussion the huge challenges of the problem cannot be understood appropriately. 
    \item Opening a discussion about the choice of existing FR-IQA measures for medical application as well as desirable properties for potential novel frameworks. They should facilitate clinical applicability and cement a more functional knowledge transfer between developers and users.
    \item Suggesting general guidelines on how to use FR-IQA in the setting of medical imaging safely, as well as highlighting an existing problem on the lack of proper reporting of employed measures.
\end{enumerate}

The idea of generalized image quality measures not being appropriate in medical imaging has been explored before, and some guidelines for task adapted quality assessment exist in the literature. While a full review is out of the scope of this work, it is worth noting the research that was started by Barret et al.~in Objective Assessment of Image Quality \cite{barrett1990objective,barrett1995objective,barrett1998objective} regarding task-based assessment with observer models, relying on the specification of a task, observer and an image ensemble. The field of objective IQA for medical imaging has been active since, in more recent works also discussing various modalities, e.g.~low dose CT~\cite{cai2017image}, MRI~\cite{mason2019comparison} or even multi-modal imaging \cite{clarkson2008task}. Lastly, on a related note, human perception in medical imaging has been an active field of research, where e.g.~The Handbook of Medical Image Perception and Techniques (ed.2) \cite{handbook} is comprehensively discussing research on image perception, observer models and clinical relevance. 

\subsection{Outline}
The paper is structured as follows. Section \ref{back} contains an overview of the most commonly used/standard FR-IQA measures and their background. In Section \ref{fail}, every subsection reports the use of these FR-IQA measures in a different medical imaging modality, including a description of the corresponding reconstruction method or visualization problem and examples of failure. Finally, Section \ref{discuss} includes a discussion, suggestions for future directions and guidelines for task-informed usage of FR-IQA measures in the context of medical imaging. 

Here, working across medical imaging domains, an international collaboration was formed that includes experts of the specific imaging techniques in order to provide insights for the different image modalities described in Section \ref{fail}. With that, we were able to ensure for every example to include at least one expert working in the particular field to provide the required insights for a comprehensive analysis and task-specific judgment of the obtained image qualities. 

\section{Background}\label{back}
The mean squared error (MSE) is a common FR-IQA metric used to measure the average squared difference between a given image and the reference, i.e.~
for a given reference image $I \in \R^{N_1 \times N_2}$ and a corresponding processed version $J \in \R^{N_1 \times N_2}$, the MSE is given by the Frobenius-norm, i.e. 
\[\MSE(I,J) := \tfrac{1}{N_1\cdot N_2}\|{I-J}\|_F^2. \]

Lower MSE values indicate that the processed image values are closer to the reference values. However, it is well known that the MSE used as a measure of image quality does not correspond well to human perception, cf.~\citep*{639240,5745362,FEI2012772, LIN2011297}, and does not provide a consistent quantity regarding severeness of image degradation. Therefore, even in the cases where the computation of the root-MSE can serve as a useful quantity measure of deviation for some medical imaging modalities (e.g.~in MRI), this would not correspond to an assessment related to a perceptual measure. 
A closely related measure was introduced in the early 2000s, the so-called peak signal-to-noise ratio (PSNR), which provides a re-scaled version of the MSE:

\begin{equation}\label{PSNR}
    \PSNR(I,J):= 10\cdot \log_{10}\left(\frac{\MAX^2}{\MSE(I,J)}\right),
\end{equation}
where $\MAX$ corresponds to the maximal possible intensity value, i.e. for an $8$-bit image $\MAX = 255$. As the PSNR solely relies on the MSE, most disadvantages and problems that are known for the MSE (e.g. same value for very different degrees of degradation) do also hold true for the PSNR. 

Two decades ago, the framework of the structural similarity index (SSIM) \citep{ssim} was introduced, which relies on three comparison components: luminance, contrast and structure, and can be calculated on various batches, i.e.~local parts of an image. For simplicity, here, we call the batches again $I$ and $J$, and then the measure is given by 
\begin{equation}
\SSIM(I,J) = \frac{(2\mu_I\mu_J + c_1)(2\sigma_{IJ} + c_2)}{(\mu_I^2 + \mu_J^2 + c_1)(\sigma_I^2 + \sigma_J^2 + c_2)},
\end{equation}
where $\mu$ corresponds to the mean, $\sigma$ to the covariance and $c_1, c_2$ are scaling factors. The values of SSIM theoretically range between -1 and 1, where a higher value indicates greater similarity between the images, in the sense that they are more visually alike. This measure has celebrated great success for natural images since the general framework allows for a greater insight if used appropriately. However, in practice, standard implementations only provide the mean over batches in the image as the final image quality measure. This choice increases the chance of failure in local error detection, which is often crucial in the medical settings. Limitations of the SSIM in the medical setting have been reported, e.g.~in \citep{limssim,ssimrad, goodpracticessim}. 
This is not surprising, since SSIM was not only not developed to assess the quality of medical images, but SSIM was also originally not tested for its use on medical images. 

Moreover, it is important to note that the SSIM framework allows to set a number of parameters, including the choice of kernel, and therefore can yield diverging results which depend on the implementation used, see the paper \citep{Venkataramanan2021AHG} for a detailed analysis of the variations. This means, for comparability and reproducibility, it should always be stated in detail which implementation/parameter setting of SSIM was used when applied. 

Many other structural FR-IQA measures have been developed in the last decade, for example the HaarPSI \citep{Reisenhofer18} measure based on Haar wavelets. Most recently, LPIPS \citep{lpips}, a highly successful FR-IQA measure for natural images, that quantifies the perceptual similarity between two images based on features learned from deep convolutional neural networks has been suggested to be included in the evaluation for medical images, see e.g.~\citep{lpipsrecom} or \citep*{jimaging9030069, CTlpips}. Although LPIPS holds some properties that are beneficial in the medical setting, including the invariance to small spatial perturbations, it has not been rigorously tested nor developed for medical images. To gain more insights of that recent development, we are including the results of LPIPS (based on the default AlexNet) applied to the provided examples, where a smaller value indicates better similarity. It is important to note that the framework generally allows the development of your own learned quality measure and provides different measures based on several networks. 

Although there is an enormous number of newly introduced FR-IQA measures - including methods developed for medical imaging tasks, see e.g.~\citep{mriiqachow}, and a published review from 2016 describing the current situation in medical imaging \citep{CHOW2016145} - the most used FR-IQA measures for assessment of classical medical imaging algorithms, such as image reconstruction, image restoration, super resolution or denoising/artefact removal, are still PSNR and SSIM. Summarizing the problem, commonly used FR-IQA measures have been developed for natural images which consist of very different properties than medical images.

\subsection{Illustration of the failure of common FR-IQA measures on synthetic image degradations}
\begin{figure}[ht!]
\centering 
    \subfigure[Reference $I$]{\includegraphics[width=0.25\textwidth]{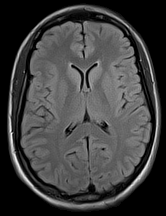}} \quad
    \subfigure[(22.6, 0.97, 0.01)]{\includegraphics[width=0.25\textwidth]{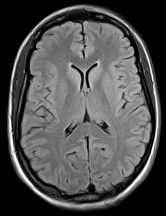}} \quad
    \subfigure[(22.6, 0.92, 0.01)]{\includegraphics[width=0.25\textwidth]{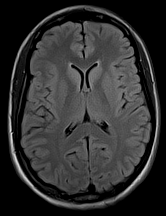}} \quad
    \subfigure[(22.6, 0.98, 0.08)]{\includegraphics[width=0.25\textwidth]{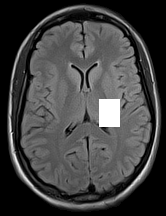}} \quad 
    \subfigure[(22.6, 0.64, 0.39)]{\includegraphics[width=0.25\textwidth]{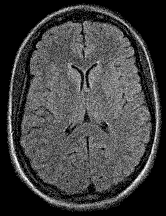}} \quad
    \subfigure[(22.6, 0.63, 0.27)]{\includegraphics[width=0.25\textwidth]{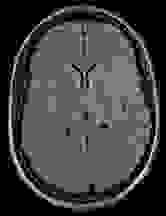}}
    \caption{Illustrative toy example of problems occuring when using the standard FR-IQA measures PSNR/SSIM/LPIPS for the evaluation of medical images. Degradations have been added artificially to the reference (a) MRI scan: contrast enhancement (b), brightness change (c), hole (d), Gaussian White noise (e), jpeg compression (f). PSNR yields the same value for all degradations, SSIM and LPIPS fail to identify the hole (d), and misjudge the quality of (e) and (f).}
    \label{toyexp}
    \end{figure}

In Figure \ref{toyexp} we show a toy example of misjudgments that occur when evaluating the quality of a 2D MRI scan compared to degraded versions of the same image with the selected measures. PSNR even yields the same value for all the very different degradations, and all of the tested measures fail in the judgement of massive local information loss (d), as well as in the judgment of stochastic noise (e) versus block artefacts (f). This toy example served as an inspiration to study the behavior of the standard measures in real-life medical imaging tasks.

\section{Examples of failure in medical imaging}\label{fail}
In this section, we will present failures of PSNR/SSIM/LPIPS when applied to medical problems with real-world image data. This contains examples of tasks where the measures are currently used as standard choices in the evaluation, as well as tasks where automated objective evaluation is still an open field and urgently needed.

The structure will be as follows. In each subsection, a medical imaging modality will be shortly introduced, followed by a problem formulation in which IQA measures play an important role. Finally, a corresponding example is shown visually with a short discussion in which regards the measures act inaccurate for that problem. 

The numbers provided in the subfigures correspond to the PSNR, SSIM, and LPIPS values in comparison to the reference image (a), respectively, where in-built functions of MatlabR2023a were used to compute PSNR and SSIM (namely, default settings for \textit{psnr} and \textit{ssim}), and for LPIPS the Python implementation based on AlexNet provided by the authors was used \citep{lpips,lpipsgithub}. We computed all measures on the visualized 2D images as shown in the paper, i.e.~no further contrast/luminance adjustment was added, to ensure visual comparability to the provided numbers. The image data was first scaled according to standards in the different fields; for the CT and photoacoustic data clipping to a pre-defined range was applied, and afterwards all images have been standardized by scaling them between $0$ and $1$ (division by the maximum, which corresponds to the maximum of the reference image in case clipping was applied). All employed image data, besides the data from digital pathology, is originally grey-scale. Depending on the original data format, the images were saved as uint8 or uint16 images in portable network graphic (png) format. \\

\subsection{Computed Tomography}
Computed Tomography (CT) 
is an imaging modality that aims to reconstruct 2D or 3D volumes from X-ray attenuation measurements and enables high-quality structural imaging of patients \citep{CT}. 
The applications of CT range from diagnostics, surgery planning and radiation therapy to image-guided interventions, making this imaging modality ubiquitous in modern medicine. 
\subsubsection{Reconstruction problem}
CT is not a direct measurement method and images need to be reconstructed by solving a large-scale system of linear equations. One of the main challenges with this task is ill-posedness, which means that in some scenarios small perturbations on the measurements can generate large perturbations on the recovered image. Particularly problematic are limited datasets, e.g. when only limited angle or sparse full-angle tomography measurements are available, as well as the presence of noise in the measurements. In these cases, the direct and most used approach to compute a solution, i.e.~the so-called filtered backprojection (FBP), can be highly corrupted by noise \citep*{CTit}. 

Different families of iterative solvers have been developed to solve a neighbouring problem that is more robust to perturbations on the measurements, see e.g.\citep{KaltenbacherNeubauerScherzer+2008,biguri2016tigre}. These iterative algorithms solve an optimization problem, refining the solution as they progress and allowing to incorporate prior knowledge via so-called regularization. However, different regularization techniques intrinsically rely on different assumptions on the reconstructed object (e.g.~smoothness or appearance of edges), which has a direct impact on the resulting quality. 

On top of that, screening requires scanning large portions of the population with harmful radiation. Therefore, taking less measurements while preserving image quality would be desirable. 
Classical regularization algorithms have been enhanced using data-driven methods, where some of the reconstruction steps are replaced by machine learning models. While these methods have a high success rate in perceived image quality, cf.~\citep{8271999, mukherjee2021learned}, the explainability is quite low. Thereto, and to increase applicability, task-adapted reconstruction for inverse problems has been introduced \textcolor{black}{into the modern data-driven pipelines}, cf.~\citep{taskad}. 

In addition to the described choice of reconstruction algorithm, the image acquisition settings (e.g.~mAs and kV) as well as the geometry parameters (e.g.~slice thickness) also influence the image quality of the CT reconstructions. 

Te following three experiments relating to the use of IQA measures in CT are presented: one on the evaluation of Krylov subspace algorithms for cone beam CT (CBCT) reconstruction, another on the evaluation of data-driven methods in lung CT reconstructions, and lastly an example on output deviations with adjusted scanner parameter settings. 

\subsubsection{Example 1: Krylov methods in CBCT}
\begin{figure}
\centering
    \subfigure[Ground Truth]{\includegraphics[width=0.2\textwidth]{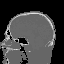}} 
    \subfigure[(23.2, 0.87, 0.12)]{\includegraphics[width=0.2\textwidth]{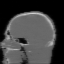}} 
    \subfigure[(28.8, 0.91, 0.03)]{\includegraphics[width=0.2\textwidth]{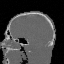}} 
    \subfigure[(28.5, 0.92, 0.03)]{\includegraphics[width=0.2\textwidth]{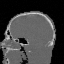}} 
    \subfigure[(25.8, 0.91, 0.05)]{\includegraphics[width=0.2\textwidth]{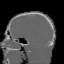}} 
    \subfigure[(28.4, 0.92, 0.02)]{\includegraphics[width=0.2\textwidth]{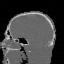}}  
    \subfigure[(28.3, 0.87, 0.03)]{\includegraphics[width=0.2\textwidth]{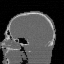}} 
    \subfigure[(28.1, 0.85, 0.04)]{\includegraphics[width=0.2\textwidth]{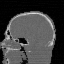}}
  \caption{CBCT Reconstructions from different Krylov methods (b)-(h) of phantom head data, and PSNR/SSIM/LPIPS values compared to the ground truth (a). The overall visual appearance is misjudged here by all three measures, e.g. PSNR in (g), SSIM in (e) and LPIPS in (g).}
  \label{fig:CBCT}
\end{figure}

The example presented here is taken from a study using Krylov subspace methods, a family of iterative reconstruction algorithms on CBCT data \citep{sabate2022krylov}. The study proposed and compared a variety of reconstruction algorithms in simulated and real CBCT problems. Here, we include an experiment involving simulated CBCT acquisitions of a head, where a mixture of Poisson and Gaussian noise is added to the measurement, to simulate realistic noise, cf.~\citep{4840547}. The performance of several Krylov algorithms was determined by comparing the final reconstructions to the ground truth, cf.~Figure \ref{fig:CBCT}. 

\subsubsection*{FR-IQA mismatches}
The reconstructions in Figure \ref{fig:CBCT} (g)(h) contain pixel-wise noise and some undesired stripe artifacts in the lower section of the head, which is not unexpected for reconstructions based on ABBA-GMRES, cf.~\citep*{hansen2022gmres}. In comparison, the other methods do not produce such artifacts, and do consist of a more uniform tissue value. However, in Figure \ref{fig:CBCT}, we can see that the computed IQA values do not penalize the loss of detailed information, and in fact PSNR/LPIPS suggest that the reconstruction in (e) is significantly worse than the ABBA-GMRES methods (g)(h), which contradicts the visual perception in these regards. SSIM on the other hand struggles here to penalize blur strong enough and gives the low-quality image in (b) a higher rating than (h). 

Quantitative assessment of novel CBCT reconstruction methods is highly needed and also encouraged to be reported when publishing a novel method. In this example, we can see that the suggested measures do not yield consistent results, and more complex image quality metrics would be required to capture both local and non-local effects appropriately.  

\subsubsection{Example 2: data driven reconstruction methods in lung CT screening}
There is sufficient evidence that screening for certain tumours using CT images may improve prognosis of cancer survivability \citep{CTlow}. As mentioned above, in order to gain better image quality with less X-ray dose, many enhanced regularization techniques with integrated machine learning steps have been suggested for CT reconstruction, and in a full reference setting they are commonly evaluated by applying PSNR and SSIM, see e.g.~\citep*{8271999, 9178467, 9433944}. As CT images are generally taken to perform a clinical task, they are not the final step of a medical process but often the initial one. Therefore the definition of what makes a good image heavily
depends on the task in hand, and for prognosis related cancer the identification of tumours is of upmost importance.  

\begin{figure}
\centering 
   \subfigure[Reference]{\includegraphics[width=0.3\textwidth]{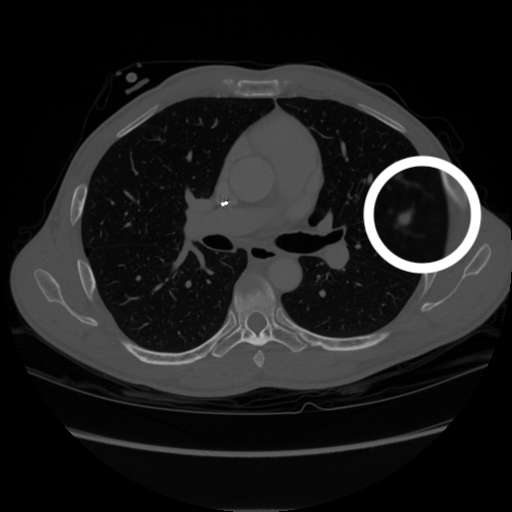}} \quad
   \subfigure[(27.6, 0.70, 0.37)] {\includegraphics[width=0.3\textwidth]{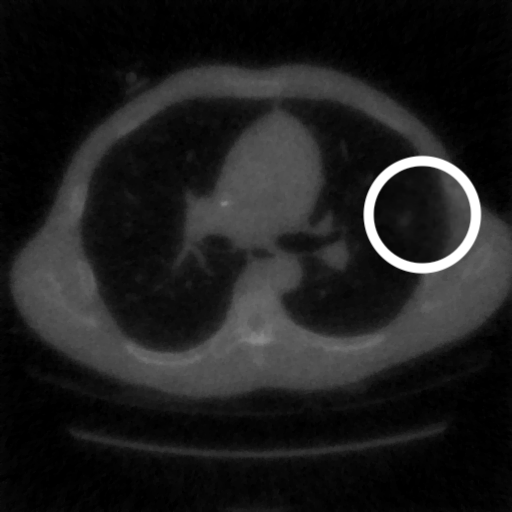}} \quad
   \subfigure[(31.9, 0.73, 0.29)]{\includegraphics[width=0.3\textwidth]{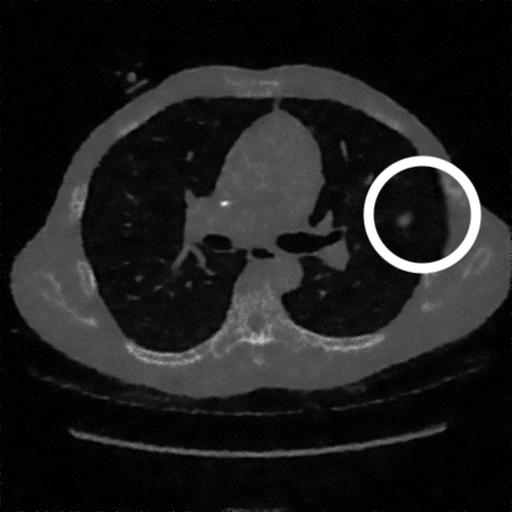}} \\
   \subfigure[(32.5, 0.84, 0.19)]{\includegraphics[width=0.3\textwidth]{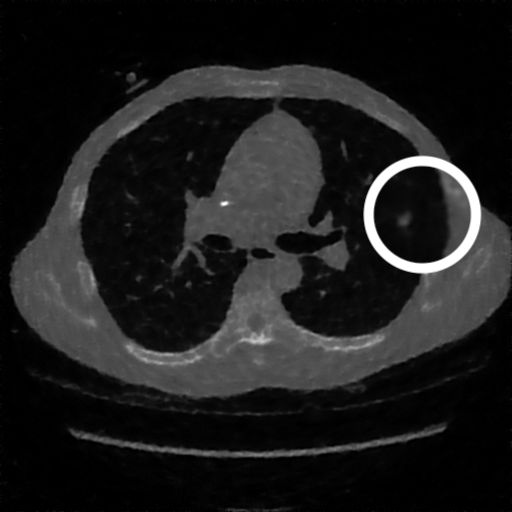}} \quad
    \subfigure[(32.5,0.77, 0.20)]{\includegraphics[width=0.3\textwidth]{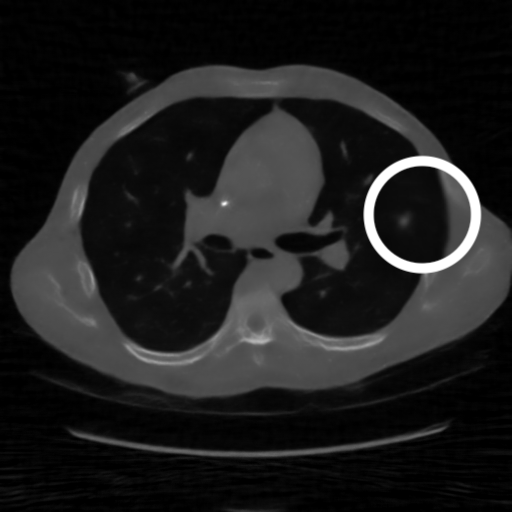}} \quad
   \subfigure[(33.0, 0.89, 0.12)]{\includegraphics[width=0.3\textwidth]{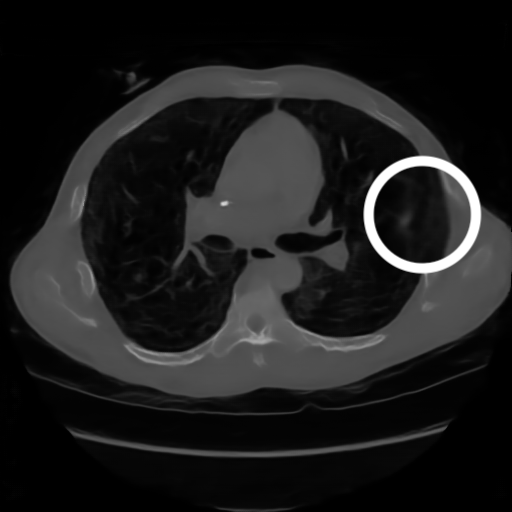}} 
    \caption{Reference image (a) and outputs of different reconstruction methods (b)-(f) applied to dose simulated data. PSNR/SSIM/LPIPS are unable to identify the best reconstruction (c), where also the tumour is visualized well.}
    \label{CT1}
\end{figure}

In ongoing research on photon counting detector types and screening procedures for lung cancer (EPSCR grant: EP/W004445/1) an experiment was conducted testing enhanced reconstruction algorithms. Simulations using less than 10\% of a clinical X-ray dose were performed to investigate if data-driven methods could sufficiently enhance the images to clearly see the tumours in the lungs while providing a very low amount of dosage to the patients. The corresponding data was a CT-dose simulation, using images from the open LIDC-IDRI dataset \citep{CTdataset} as references, as well as simulated and reconstructed images with in-house software. Figure \ref{CT1} shows the results of the experiment. We show the reference image used as a basis for the simulation, together with five different reconstruction algorithms. The first is an iterative solver, a gradient descend algorithm with TV minimization \citep{champock} and (c)-(f) correspond to machine learning methods: FBPConvnet is a denoising algorithm that cleans the bad image \citep{7949028}, LPD is an iterative unrolled method that combines traditional solvers with machine learning \citep{8271999}, Noise2Inverse is a self-supervised learning method (i.e.~does not require ground truth data) \citep{9178467} and ItNet is another iterative unrolled method, the best performing winner of the AAPM DL-Sparse-View CT challenge \citep*{genzel2021aapm}. ItNet is also judged here as the best result according to PSNR, SSIM and LPIPS.  

\subsubsection*{FR-IQA mismatches}
This experiment was performed to evaluate the quality of different kinds of CT reconstruction, especially the lung tumour detection capabilities thereof. The best result according to the chosen IQA measures is given by ItNet in Figure \ref{CT1}(f), which performs visually poorly. Not only the tumour (zoomed in white circle) is significantly less visible in the reconstruction, but ItNet also produces structures in the lung that are different than the ones in the reference image; it blurs and lengthens much of the soft tissue present in the lungs and it also created structure from noise in some places. Moreover, the image is overly smooth. Comparing the other reconstruction algorithms, it seems that FBPConvnet Figure \ref{CT1}(c) is the one performing best at preserving the shape of the lung nodule, even when the resulting image contains enhanced pixel-level noise. 

We can see here that the qualitative findings strongly contradict the numbers provided by the selected measures. The reconstruction of ItNet, Figure \ref{CT1}(f), outperforms the other reconstructions in regard to the measures, and the qualitative winner FBPConvnet, Figure \ref{CT1}(c), is judged as second worst by the same measures. This experiment suggests that the discussed measures are not a good choice for that kind of CT reconstruction application and are yielding misleading results. 

While pixel-independent random noise may be a worse effect in a natural image than a slightly oversmooth reconstruction, this is not true in CT images, where small structures may disappear if smoothing is promoted against edge preservation. In iterative reconstruction algorithms, such choices are explicitly made by choosing the prior appropriately, and in data-driven models the researcher has limited control on the type of implicit priors the algorithm learns from the data, i.e. model builders do not know what the algorithms choose to learn from the ground truth. In these cases, appropriate evaluation would be even more important to ensure the described quality properties. 

\subsubsection{Example 3: Scanner settings impact in IQ}
Changing CT scanner settings, like tube voltage or reconstruction geometry, has a direct impact in the noise distribution of the data and thus in the quality of the reconstructed images. Here, we show an example of quality differences with acquired CT data from a realistic silicone phantom fabricated with multi-material extrusion 3D printing technology \citep{HATAMIKIA2023}. The phantom model was derived from an abdominal CT and was fabricated with realistic radio density values which could mimic imaging properties of soft tissues in CT.  

For the reference image, the anatomical phantom was scanned with the 
standard clinical CT protocol from SOMATOM Definition AS scanner, Siemens Healthineers, Erlangen Germany (tube current time product 70mAs for samples and 150mAs for anatomical phantom, tube voltage 120kVp, slice thickness 0.60mm, pixel spacing 0.77mm, iterative reconstruction kernel J30s). Additional scans with varying kVp values (80/100/120) as well as varying slice thickness (0.6/2mm) were also performed to assess the effect of the  parameters on the image quality. We observed that changing kVp and slice thickness resulted in different image quality, where higher kVp and smaller slice thickness give the best visual result.

\begin{figure}
\centering
    \subfigure[Reference]{\includegraphics[width=0.3\textwidth]{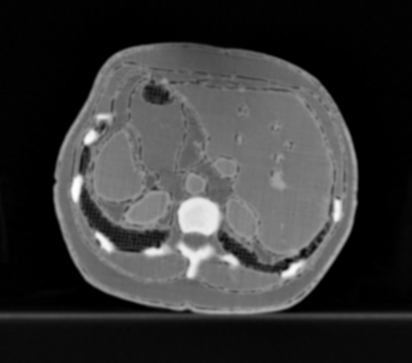}}
     \subfigure[(26.3, 0.88, 0.07)]{\includegraphics[width=0.3\textwidth]{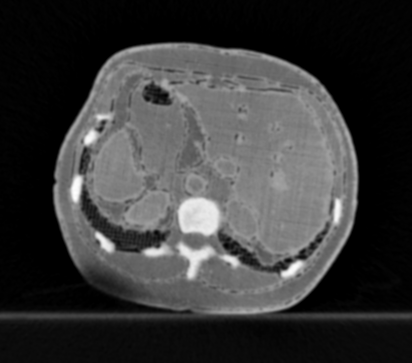}}
      \subfigure[(28.2, 0.95, 0.03)]{\includegraphics[width=0.3\textwidth]{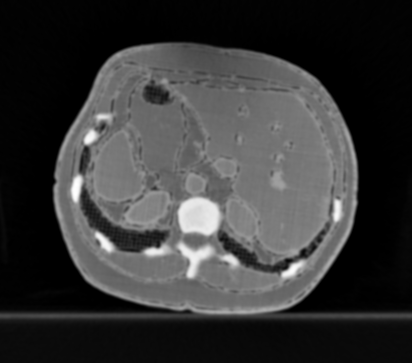}} 
    \caption{Comparison of image acquisition settings, (a) reference image with best chosen parameter setting (0.6mm and 120kVp), (b) preserves more detail (0.6mm and 80kVp) than (c) which is more smoothed (2mm and 100kVp). PSNR/SSIM misjudge the visual quality, LPIPS yields reasonable quality scores here.}
    \label{CT2}
\end{figure}

\subsubsection*{FR-IQA mismatches}
Although all IQA measures yield a better value for the image shown in Figure \ref{CT2}(c), a higher visual correspondence with the reference image can be seen in Figure \ref{CT2}(b) despite the black shadow in the bottom left corner. The image in Figure \ref{CT2}(c) with lower kVp yields a result that is too smooth in comparison to the reference. This yields another CT example where the IQA measures have been misled by quality properties that are not relevant for the clinical application. 

\subsection{MRI}
Magnetic Resonance Imaging (MRI) is a non-invasive medical imaging modality that provides excellent image quality tissue structure without ionizing radiation, but on the other hand is relatively slow. The acquired 3D data, sampled in the k-space domain, corresponds to the Fourier transform of the spatial-domain MR image. To reconstruct an accurate MR image, sampling theory indicates the total number of k-space data that must be acquired to avoid artifacts in the reconstruction. As this number is relatively large and cannot be arbitrarily reduced, shortening the total scan time compromises the image quality  \citep{mriscantime,https://doi.org/10.1002/pamm.202000159}.

\subsubsection{Reconstruction Problem}
MRI requires long acquisition times, directly related to the final resolution and tissue contrast. For many clinical applications faster data acquisition is necessary to minimize the stress on the patient and, moreover, it is important to reduce physiological motion as much as possible since this causes artifacts in the images. In order to fasten acquisition, but still receive reasonable image quality, several approaches have been introduced, see \citep{https://doi.org/10.1002/jmri.28462}. Most of these techniques acquire less data than theoretically required.
To avoid low quality due to less sampled data, techniques such as parallel imaging \citep{sense,grappa} and compressed sensing \citep{4472246}, have been successfully employed in the past decades. More recently, aiming for even more advancement, machine learning methods have demonstrated promising results. The goal is to achieve a high acceleration factor while preserving the imaging quality. The acceleration factor is given by the ratio of the amount of k-space data required for a fully sampled image to the amount collected in an accelerated acquisition. The outputs of such methods are usually evaluated with PSNR and SSIM, see e.g. \citep{10.1007/978-3-031-43999-5_47,ZHOU2022102538}.  

\subsubsection{Example 1: Scan acceleration}
For this example the data is obtained from the publicly available fastMRI brain dataset \citep{fastmriall}, which consists in total of $6405$ T1, T2 and FLAIR 3D k-space volumes. The fastMRI challenge series provided MRI datasets to foster the development of accelerated reconstruction algorithms. The series consists of a knee MRI dataset and challenge in 2019 \citep{fastmriknee}, of a brain dataset and challenge in 2020 \citep{fastmribrain}, and of a prostate dataset in 2023 \citep{fastmriProstate}. The winners of the challenges were selected by the comparison of the provided reference images, created by the rSOS of the fully sampled data, to the image outputs of the proposed method via the SSIM and the highest ranked results were submitted to receive experts' opinions. 

\begin{figure}[ht!]
\centering
      \subfigure[(27.8, 0.78, 0.14)]{\includegraphics[angle=180,origin=c,width=0.26\textwidth]{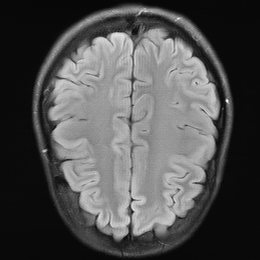}} \ 
      \subfigure[(29.6, 0.85, 0.14)]{\includegraphics[angle=180,origin=c,width=0.26\textwidth]{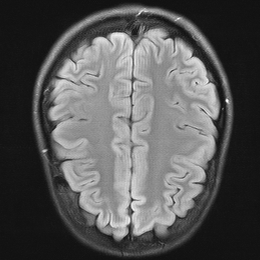}} \ 
    \subfigure[(32.3, 0.92, 0.04)]{\includegraphics[angle=180,origin=c,width=0.26\textwidth]{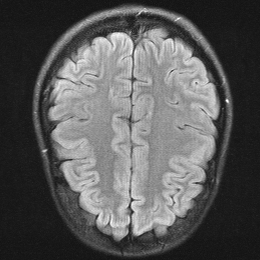}}    
            \\    
    \subfigure[(29.5, 0.84, 0.12)]
      {\includegraphics[angle=180,origin=c,width=0.26\textwidth]{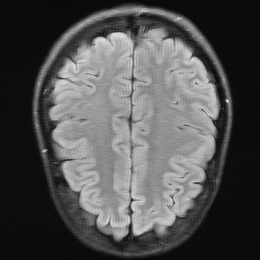}} \ 
     \subfigure[(29.6, 0.86, 0.12)]{\includegraphics[angle=180,origin=c,width=0.26\textwidth]{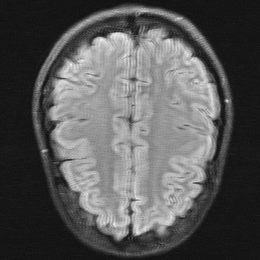}} \ 
     \subfigure[(33.0, 0.94, 0.04)]{\includegraphics[angle=180,origin=c,width=0.26\textwidth]{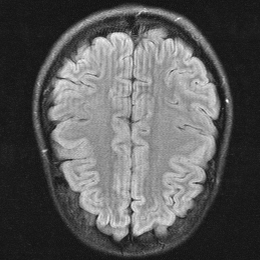}} 

    \caption{Reconstruction outputs of accelerated FLAIR MRI data from the algorithms \textit{Xpdnet}(a)(d) and \textit{E2varnet} (b)(c)(e)(f). The bottom images (d)-(f) are judged by PSNR/SSIM/LPIPS as better reconstructions than the respective image above them (a)-(c), although they contain stronger blur and contain more ringing artifacts.}
    \label{mri}
\end{figure}

We show here images obtained from two machine learning reconstruction algorithms that took part in the fastMRI multi-coil brain dataset challenge in 2020, namely the end-to-end variational network \textit{E2E-VarNet} \citep{e2evarnet} and \textit{XPDNet} \citep{xpdnet}. \textit{XPDNet} was among the top three submissions of the challenge and both algorithms perform very well on the corresponding public leaderboard \citep{leaderboardfastmri}, that allows comparison of algorithms submitted after the challenge deadline. The authors of the \textit{XPDNet} algorithm provided two distinct models for different acceleration factors. Here, we employ the neural network provided for acceleration factor 4. The reconstructions in Figure \ref{mri} were obtained by the application of \textit{E2E-VarNet}, Figure \ref{mri} (b)(c)(e)(f), and \textit{XPDNet}, Figure \ref{mri} (a)(d), to sub-sampled data with random masks (acceleration factor between 1 to 5) in the frequency domain. 

\subsubsection*{FR-IQA mismatch}
We can see in Figure \ref{mri} that the visual quality of the obtained images does not correspond to the numbers provided by PSNR/SSIM/LPIPS, since the images with better numbers (bottom row) suffer from information loss due to blur and ringing. This is not surprising as some challenges with SSIM as a performance metric have already been discussed and shown in the official results paper of the fastMRI challenge \citep{fastmriresults}. 
Here, we complement with examples where the visual results also ask for a different judgement in a non local manner. 
Curiously, the degraded images (e) (f) do hold quite higher numbers in comparison to (a) which is nearly noise-free.

\subsubsection{Example 2: Diffusion-weighted MRI (dMRI)}
DMRI is an important MRI technique to study the neural architecture and connectivity of the brain. It is based on obtaining multiple $3$-dimensional diffusion-weighted images to investigate the water diffusivity along various directions, being clinically important especially for the investigation of brain disorders, see e.g.~\citep{diffrev}. However, low signal-to-noise ratio and acquisition time limit the spatial resolution of dMRI and therefore its usage is currently mainly restricted to medium-to-large white matter structures, whereas very small cortical or sub-cortical regions cannot be traced accurately. To overcome this, several methods for increasing the spatial resolution of dMRI have been introduced, see e.g.~\citep{DYRBY2014202, srrdiff, highresdiff}. 

Here, we study image data from an acquisition and reconstruction scheme for obtaining high spatial resolution dMRI images using multiple low resolution images, cf.~\citep{dmrihighres}. The suggested method combines the concepts of compressed sensing and super-resolution to reconstruct high.resolution diffusion data while allowing faster scan time. The data is visualized via the fractional anisotropy (FA) measures computed using diffusion tensor imaging \citep*{dti}. 

The data from a human subject was acquired from a MGH connectome 3T scanner. Three thick-slice diffusion weighted imaging (DWI) volumes with voxel size $0.9 \times 0.9 \times 2.7 mm^3$, TE/TR = $84/7600 ms$ and $60$ gradient directions at $b= 2000 s/mm^2$. A separate low-resolution isotropic DWI with a spatial resolution of $1.8 \times 1.8 \times 1.8 mm^3$ and with $60$ gradient directions at $b = 2000 s/mm^2$. 

\begin{figure}
\centering
   \subfigure[Reference]{\includegraphics[width=0.3\textwidth]{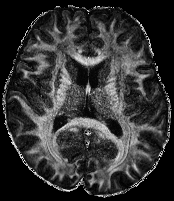}} 
      \subfigure[(17.62, 0.57, 0.19)]{\includegraphics[width=0.3\textwidth]{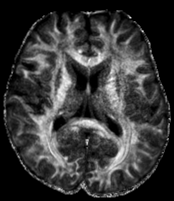}}      
      \subfigure[(18.23, 0.58, 0.26)]{\includegraphics[width=0.3\textwidth]{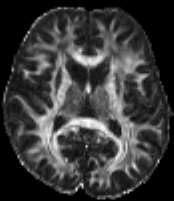}} 
    \caption{Visualized FA images obtained from diffusion MRI with super-resolution reconstructions. The up-sampled image (c) with lower resolution is wrongly judged to have better quality than the high-resolution reconstruction (b) by PSNR and SSIM, LPIPS judges this task correctly.}
    \label{mri2}
\end{figure}

The super resolution image in \ref{mri2}(a), serving as a reference here, was obtained using the super-resolution reconstruction technique that combines 
multiple thick-slice DWI with all 60 diffusion directions into a high-resolution image, cf.~\citep{dmriref}. This technique yields a high quality image with good detail preservation, but takes much longer scan time than the standard upsampling method in \ref{mri2}(c), where the FA map of the low-resolution data was up-sampled using \textit{3DSlicer} \citep{3dslicer} to the higher resolution.

The image in \ref{mri2}(b), cf.~\citep{dmrihighres}, was obtained using a combined super-resolution reconstruction, compressive sensing, and spatial regularization techniques with thick-slice images, where each thick-slice DWI has a different set of $20$ diffusion gradient directions, saving indispensable scan time. 
The advanced method yields a much higher visual quality image than \ref{mri2}(c), preserving more anatomical details.

\subsubsection*{FR-IQA mismatch}
We can see in Figure \ref{mri2} that PSNR and SSIM misjudge the visual quality of the high-resolution reconstruction in (b) in comparison to the up-sampled image in (c). The image is per default more blurry and does not provide sufficient anatomical details and therefore offers worse visual quality than the reconstruction in (b). LPIPS yields more sufficient results in this example, and correctly attributes (c) a higher quality error. 

In this example, it has to be noted that the computed IQA numbers are generally quite low, because the resulting FA images do not necessarily have the same range or distribution as the reference image. Therefore, in order to compare the reconstruction quality directly, this task generally benefits from NR-IQA evaluation.

\subsection{X-Ray}\label{xray}
X-ray imaging is a fundamental form of radiography. Reducing radiation dose while maintaining image quality is a key principle in radiology known as ALARA (As Low As Reasonably Achievable) \citep{alara}. New technologies and imaging techniques, such as post-processing by artificial intelligence (AI) \citep{HARVEY20201479}, may allow diagnostic objectives to be achieved with lower radiation doses. Furthermore, advancements in X-ray have also the potential to influence and enhance Computed Tomography (CT) \citep{FEGHALI2021463}. 
Whereas CT requires complex imaging reconstruction algorithms, X-ray is more straightforward, employing post-processing for high-quality and detailed imaging, being crucial for clinical assessment of anatomical structures and potential pathologies.

\subsubsection{Post-processing problem}
The raw data captured during digital radiography reflects the pattern of X-ray attenuation by different tissues. The digital signal is then processed to create a greyscale image, where each shade corresponds to the radiodensity of the tissues, ranging from black for air through varying shades of grey for soft tissue and white for bone. Post-processing software refines the raw image to enhance clarity and diagnostic utility \citep{SEERAM200823}. This may involve adjusting parameters such as brightness and contrast, applying filters for noise reduction, or using algorithms for edge enhancement \citep{xraydisp}. The aim is to produce an image that provides the best possible diagnostic information while adhering to the ALARA principle. 

Different quality properties may be desired depending on the purpose of the X-ray. For example, when visualizing the lung tissue, adjustments are made to the brightness and contrast to best highlight the anatomy and common abnormalities while minimizing noise. However, noise is less important when aiming to confirm the position of a line, tube, or foreign object. In this case, an image with edge enhancement and adjusted brightness levels may be desirable to amplify the distinction between the dense material of the object and the surrounding soft tissue.

Quality control for the provided default post-processing is usually made by the manufacturers themselves, therefore not accessible to the end user, and may be divergent to the IQ needed for clinical images \citep{https://doi.org/10.1002/acm2.14285}. After the machine has been placed in the hospital environment, personalized post-processing settings are often determined by subjective visual inspection. 
Objective evaluation would help to find an optimal post-processing type for visualization, allowing faster and consistent evaluation. Beyond this setting, objective IQA is important for the development and testing of machine learning algorithms on chest X-ray images including super-resolution, denoising or inpainting methods, where PSNR and SSIM are currently the standard choice for quality assessment, see e.g.~\citep{10.1145/3426020.3426088, jimaging4020034, xrayden, xrayden2, xrayden3}.

\subsubsection{Data}
\begin{figure}
  \centering
  
  \subfigure[Reference]{ \includegraphics[width=0.3\textwidth]{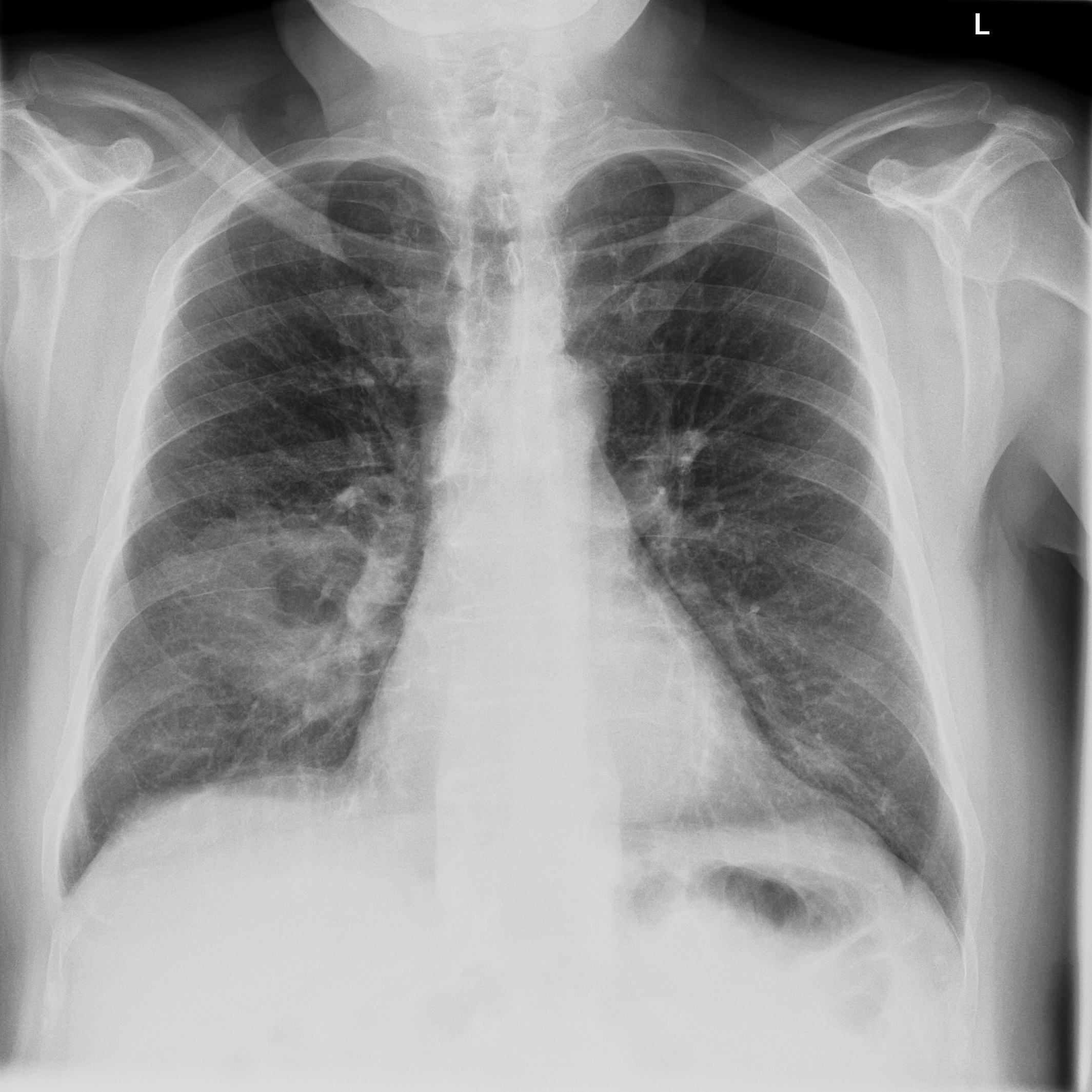} }
\subfigure[(21.1, 0.90, 0.11)]{  \includegraphics[width=0.3\textwidth]{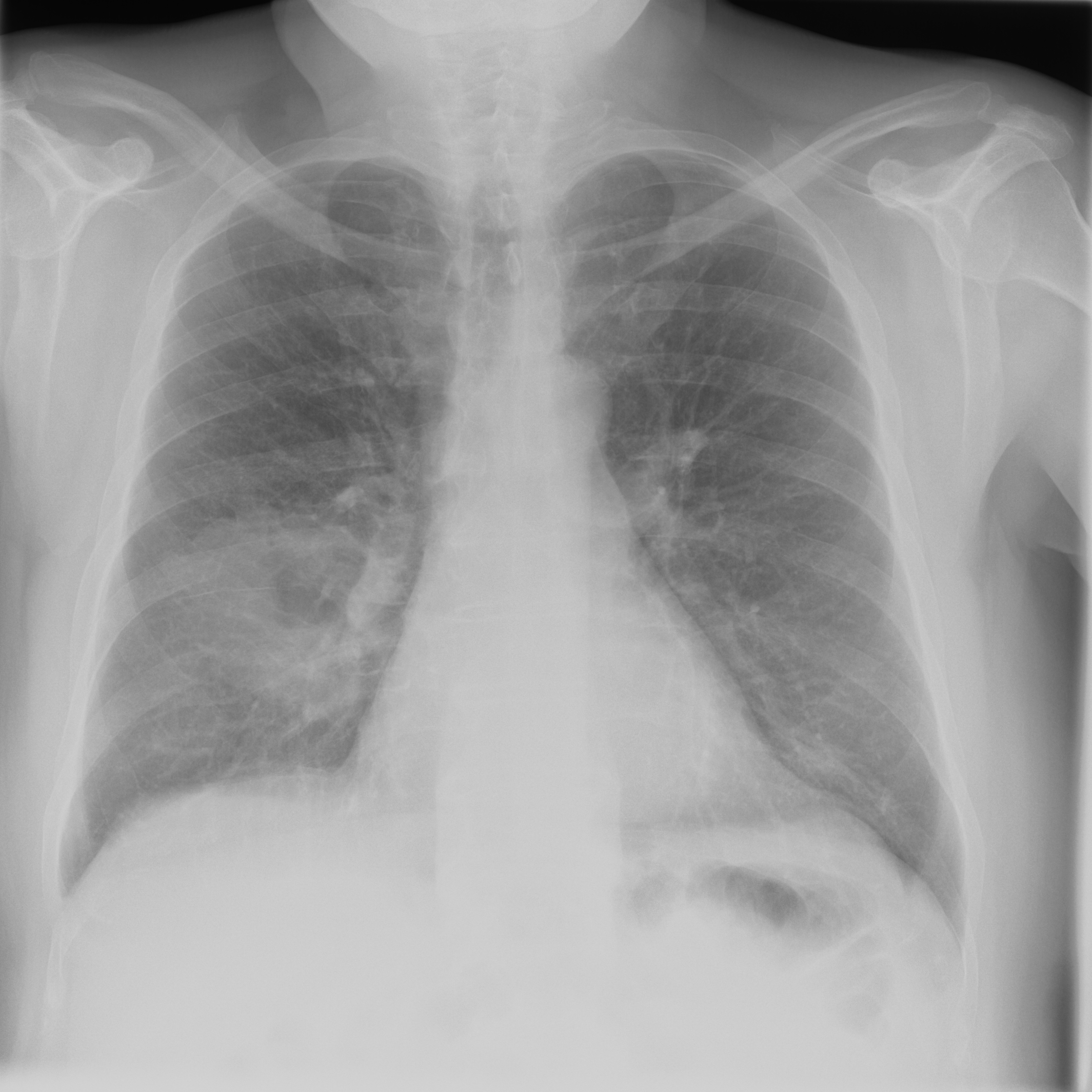}}
 \subfigure[(19.5, 0.88, 0.16)]{ \includegraphics[width=0.3\textwidth]{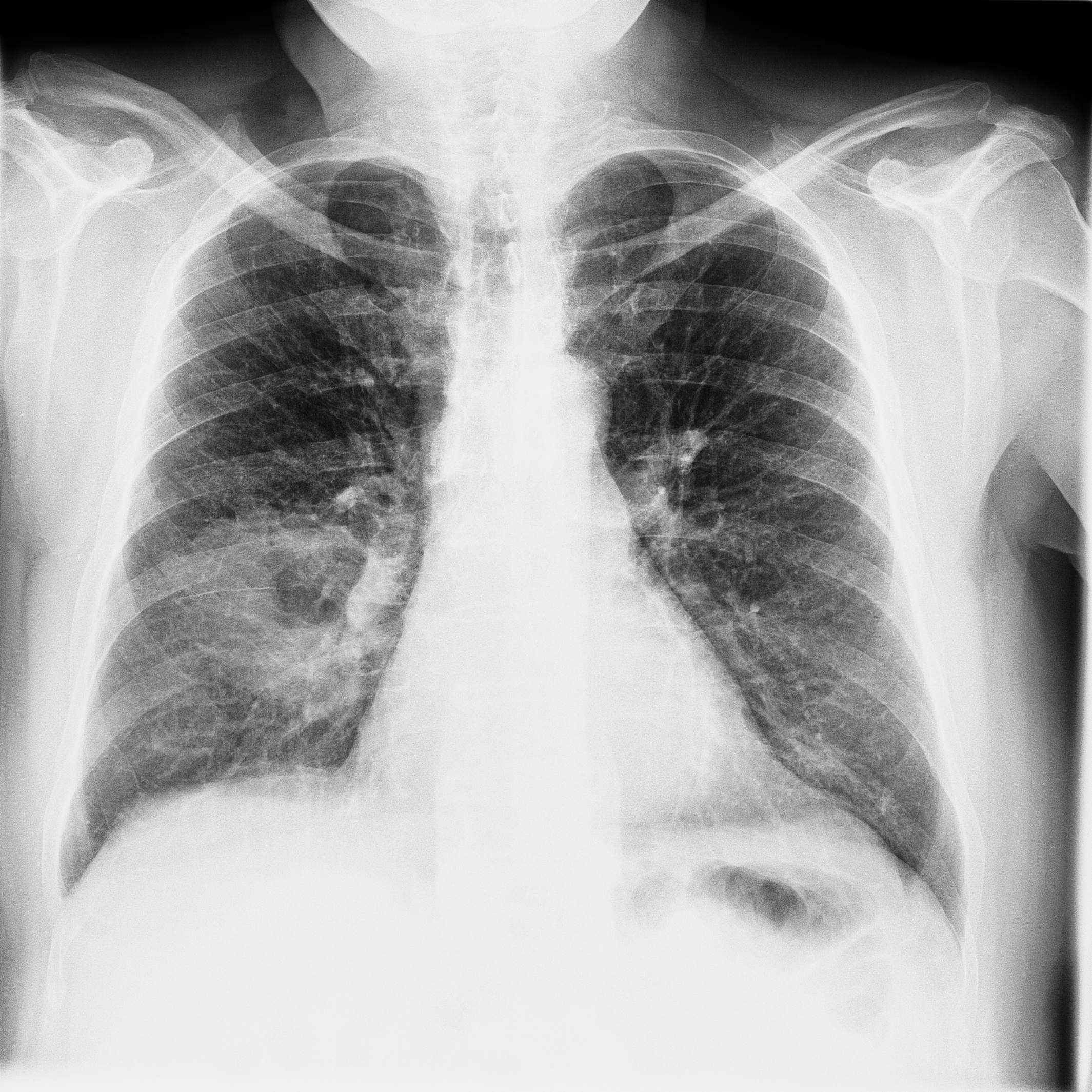}}
  \caption{Chest X-Ray scans with different kinds of post-processing; (a) serves as reference and (b) is wrongly judged as better visualization by PSNR/SSIM/LPIPS. }
  \label{xray1}
\end{figure}

\begin{figure}
  \centering
  \subfigure[Reference]{ \includegraphics[width=0.3\textwidth]{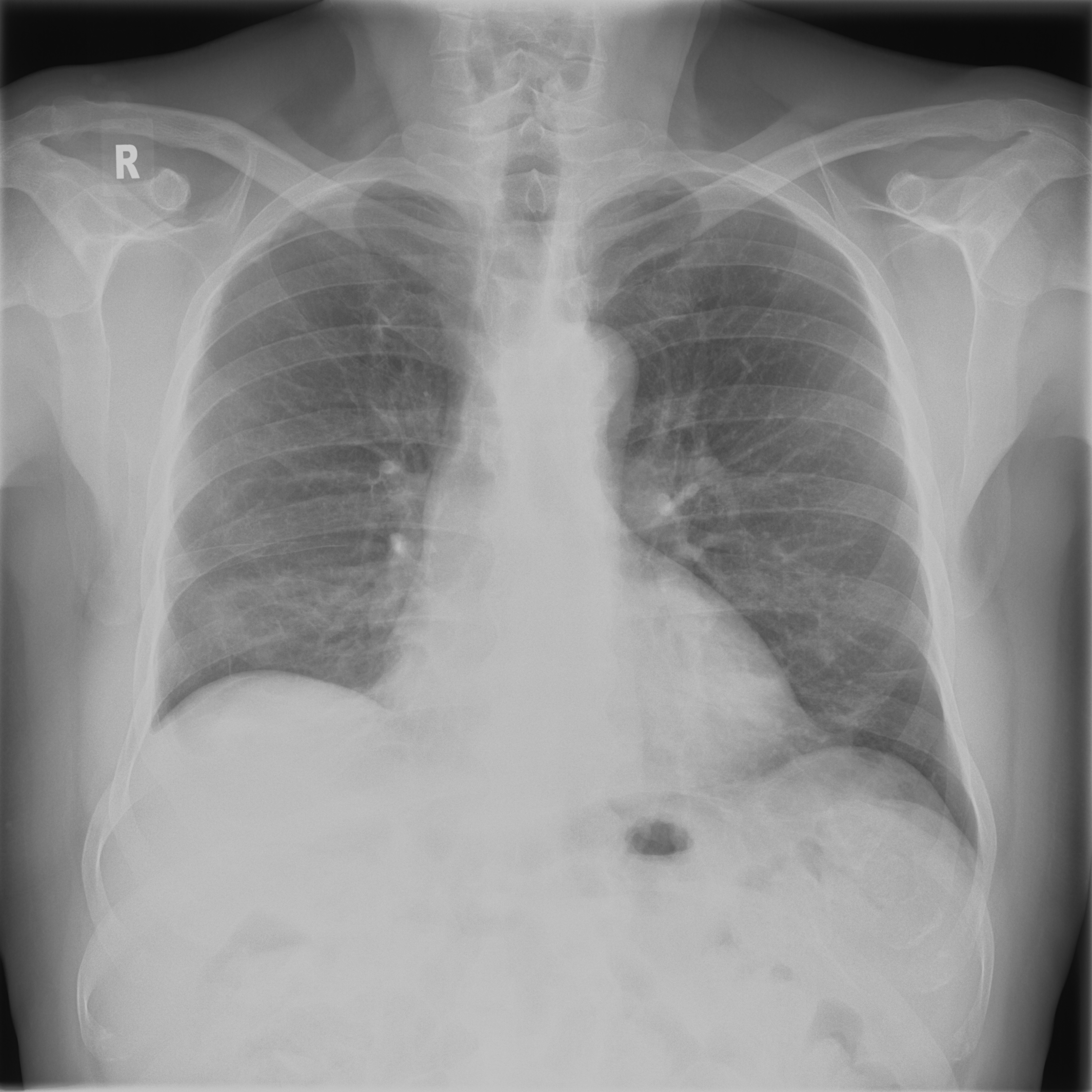} }
\subfigure[(22.8, 0.97, 0.08)]{  \includegraphics[width=0.3\textwidth]{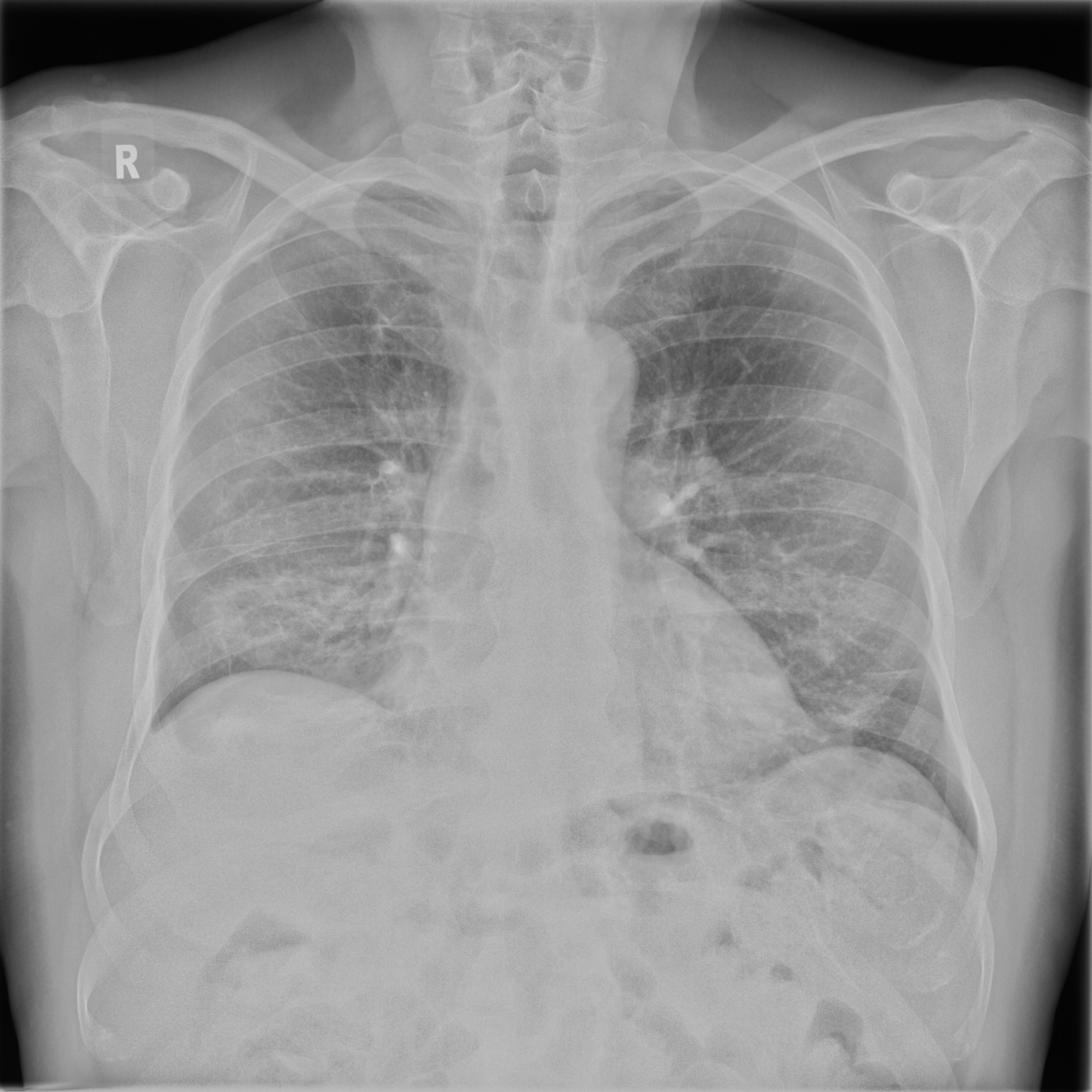}}
 \subfigure[(21.2, 0.96, 0.07)]{ \includegraphics[width=0.3\textwidth]{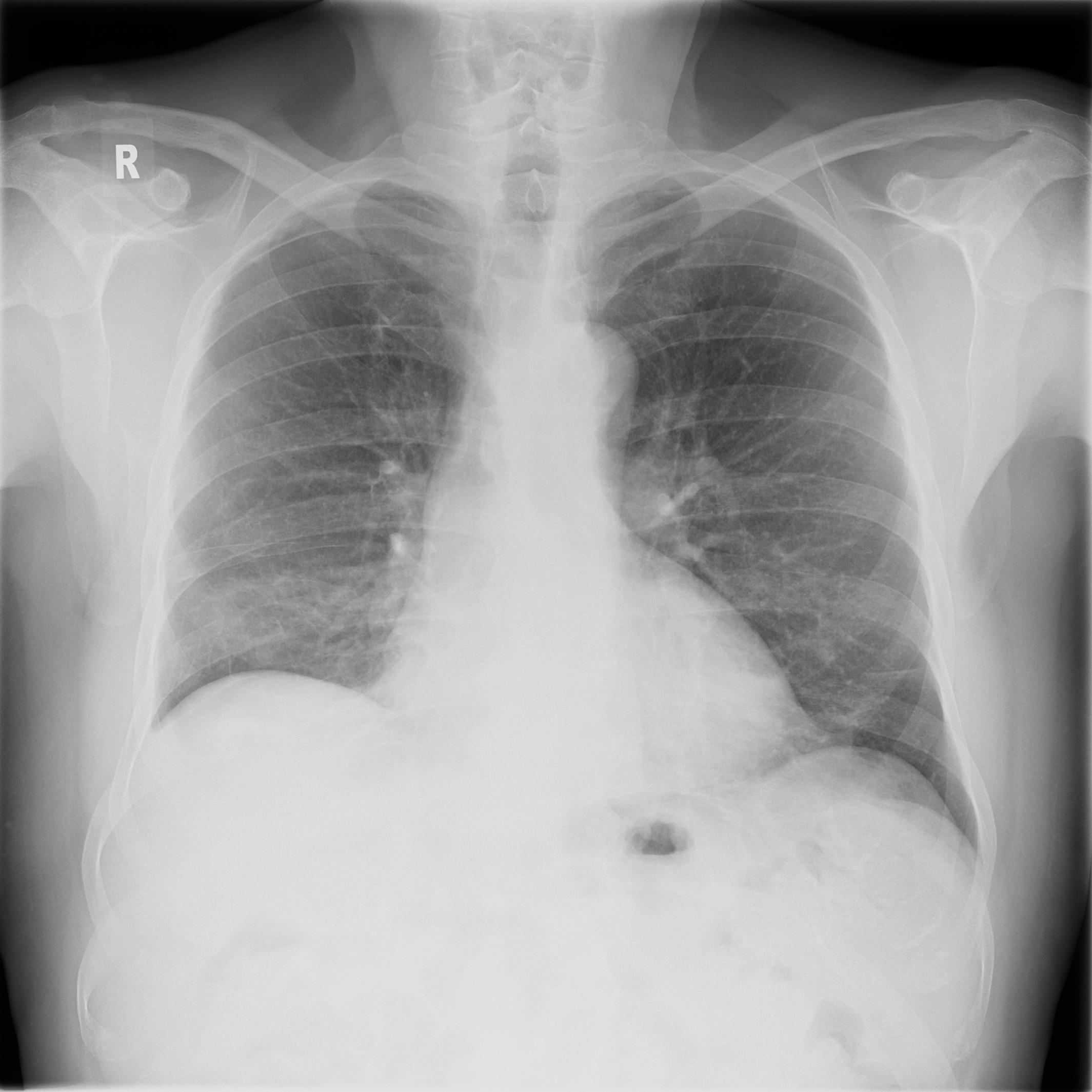}}
  \caption{Chest X-Ray scans with different kinds of post-processing; (a) serves as reference, (b) is wrongly judged as better visualization by PSNR and SSIM, LPIPS gives a slightly worse evaluation for (b).}
  \label{xray2}
\end{figure}

Posteroanterior chest radiographs were acquired on two imaging systems (both Discovery XR656 HD models, GE Healthcare, USA) at Cambridge University Hospitals NHS Trust. Each scanner was being set up in the hospital with different post-processing parameters (chosen by the operating radiologists), which are used here as reference images, see Figure \ref{xray1} and \ref{xray2}(a). Additional images, serving as real-life examples of lower quality, were produced for each radiographic exposure using multiple different post-processing settings. The post-processing was applied in the hospital directly on the scanner itself by adjusting parameters in the provided framework. 

\subsubsection*{FR-IQA mismatches}
In Figure \ref{xray1}, contrast deviation and edge enhancement were reduced in (b), but increased in (c),  the noise reduction algorithm was removed in both. The brightness was increased in both images but more so in (c) and low-contrast enhancement was removed in (b). The result is that (b) has relatively low contrast in the lungs compared to the reference (a) and radiograph (c). In Figure \ref{xray2}, edge enhancement has been dramatically increased in (b), whilst the contrast deviation and tissue contrast have been reduced. In (c), the brightness, tissue contrast and edge enhancement have been slightly increased. Consequently, (b) provides low contrast in the lungs with excessively prominent lung markings and vasculature which make it harder to detect abnormalities such as a pneumonia. 

All of the chosen FR-IQA metrics wrongly judge (b) as the better image in the first example Figure \ref{xray1}, and the results in (b) and (c) of the second example Figure \ref{xray2} are quite close, where PSNR and SSIM are also providing the wrong order. The tested measures are not suitable to evaluate the quality of data sets with X-Ray images that have large variations regarding contrast, luminance and sharpness. 

\subsection{OCT}
Optical coherence tomography (OCT) is a well established imaging technique based on low-coherence interferometry that enables volumetric imaging of biological tissue at high resolution \citep{PIRCHER2018155}. Light is split into a reference and a sample arm, recombined after being backreflected by a mirror and backscattered at different depths of the sample, in the respective paths. Using Fourier domain OCT \citep{FERCHER199543}, a tomographic image of the sample is reconstructed by spectral analysis of the resulting interference patterns. In order to achieve optimal axial resolution, it is essential to match dispersion between the two arms of the interferometer \citep{Drexler2008}. This can be achieved on the one hand by carefully matching the length of the arms and the corresponding dispersive materials and on the other hand by numerical methods  in the reconstruction process \citep{Wojtkowski:04}. OCT was originally developed for imaging of the retina and, while there are many other applications of OCT available, its highest impact is to this day in  ophthalmology, where this technology plays a critical role in correct diagnosis \citep{PIRCHER2018155}. 

\subsubsection{Reconstruction problem}
Most commonly, OCT processing algorithms compute intensity images from the recorded spectral data. Standard algorithms include a step that performs numerical dispersion compensation. Methods are available which automatically determine the correct dispersion compensation parameter \citep{Wojtkowski:04}, but the stability of automatic numerical dispersion compensation methods under varying imaging conditions is not yet fully understood. Besides the dispersion parameter, there exist also algorithms that provide geometrical corrections within the reconstruction process, 
see e.g.~\citep{Shirazi:20}, namely the correction of rotation introduced by eye motion and correction of the curvature of the retina. Automatic parameter selection for the geometric corrections is a challenging task which has so far not been addressed and therefore these parameters are usually set manually, which is very time-consuming especially in the case of clinical studies where often a large number of patients and imaging locations are included.  

\subsubsection{Data}
\begin{figure}[ht!]
  \centering
  \subfigure[Reference]{ \includegraphics[width=0.44\textwidth]{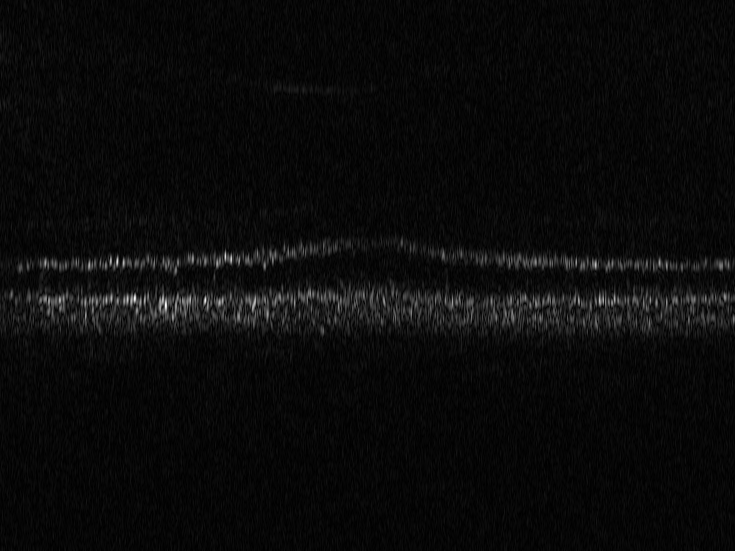}}
\subfigure[(26.88, 0.71, 0.05)]{  \includegraphics[width=0.44\textwidth]{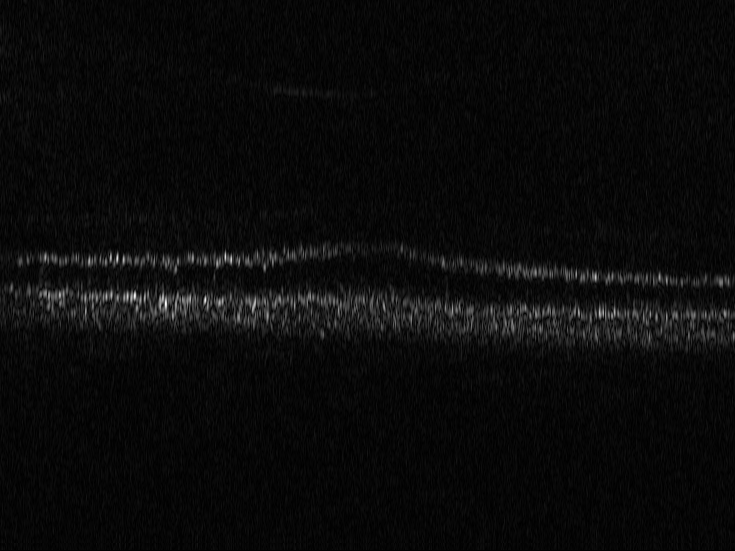}}
\subfigure[(26.84, 0.72, 0.04)]{  \includegraphics[width=0.44\textwidth]{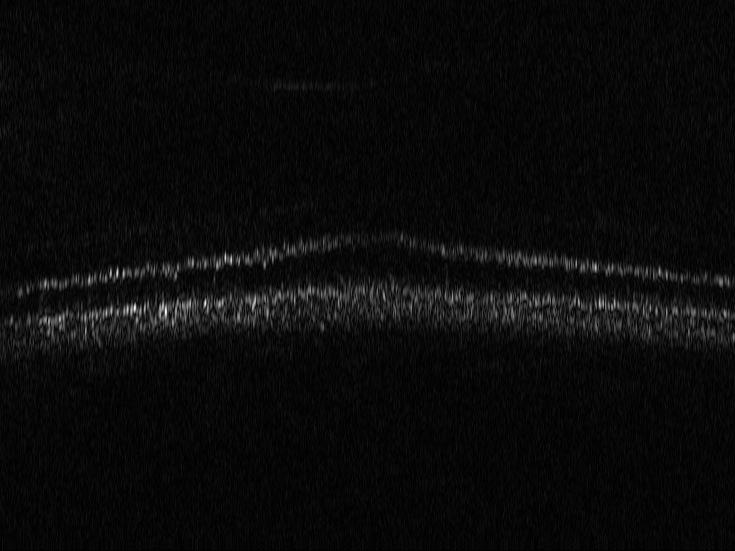}}
 \subfigure[(29.94, 0.77, 0.06)]{ \includegraphics[width=0.44\textwidth]{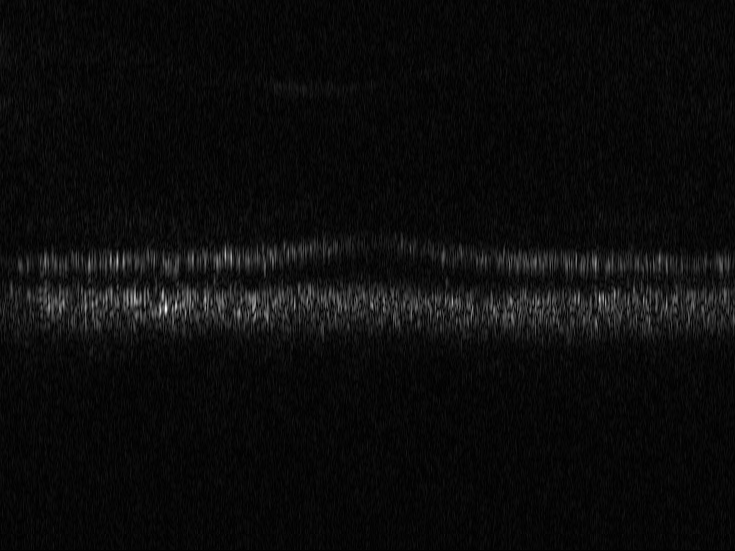}}
  \caption{OCT reference reconstruction (a) and reconstructions with sub-optimal parameters (b)-(d) leading to geometric deviations (b)-(c) and low resolution (d). Here, (d) is wrongly judged as best reconstruction by SSIM and PSNR, LPIPS is able to ignore the small spatial deviations. }
  \label{aooct1}
\end{figure}

\begin{figure}[ht!]
  \centering
  \subfigure[Reference]{ \includegraphics[width=0.44\textwidth]{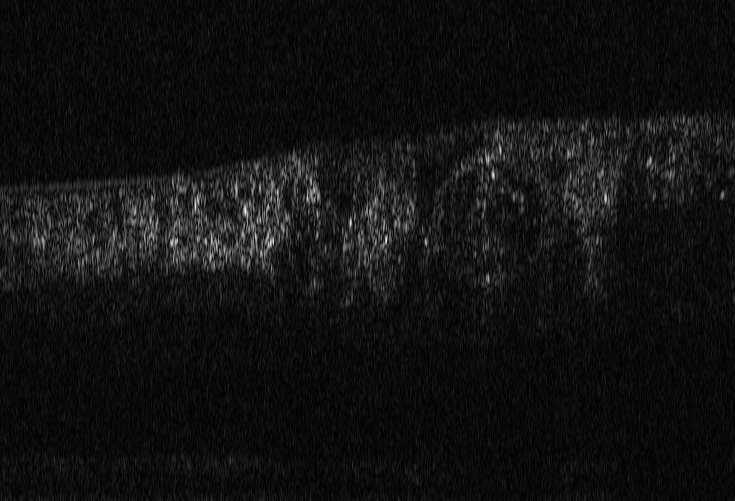}}
\subfigure[(25.20, 0.55, 0.04)]{  \includegraphics[width=0.44\textwidth]{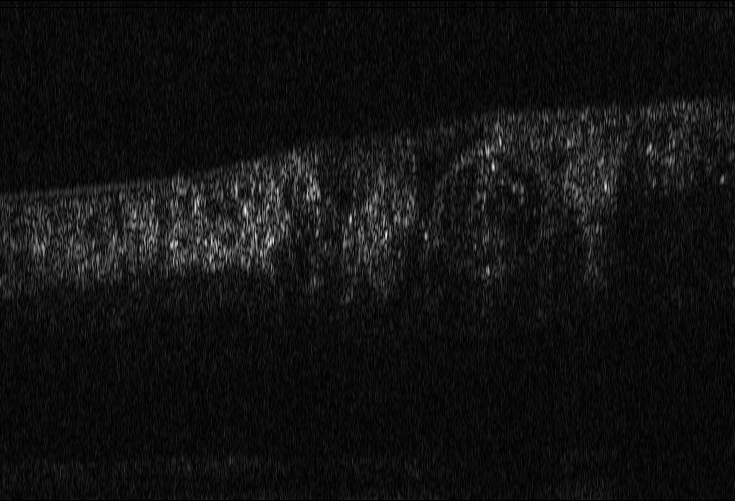}}
\subfigure[(26.49, 0.66, 0.03)]{  \includegraphics[width=0.44\textwidth]{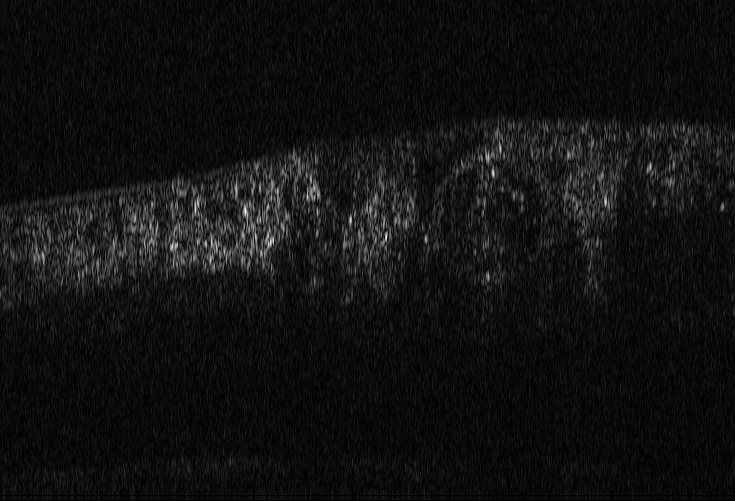}} 
 \subfigure[(27.78, 0.70, 0.06)]{ \includegraphics[width=0.44\textwidth]{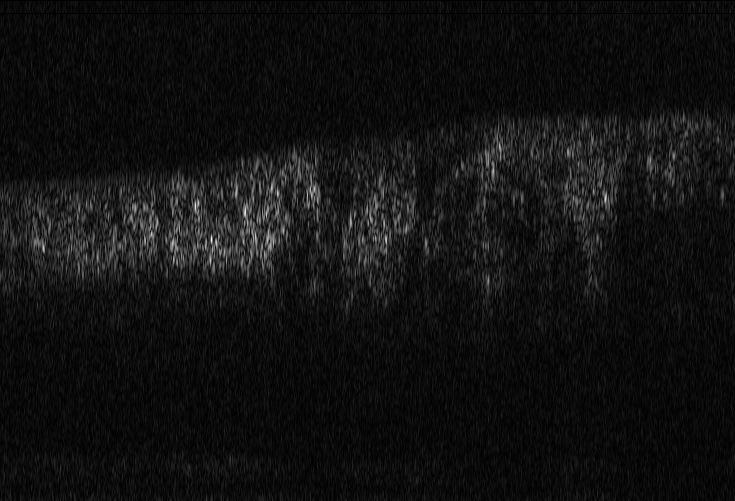}}
  \caption{OCT reference reconstruction (a) and reconstructions with sub-optimal parameters (b)-(d) leading to geometric deviations (b)-(c) and low resolution (d). Here, (d) is wrongly judged as best reconstruction by PSNR and SSIM, LPIPS is able to ignore the small spatial deviations.}
  \label{aooct2}
\end{figure}

We employ image data obtained using an adaptive optics (AO) supported spectral domain OCT system \citep{Brunner:21}, where cross-sectional images (B-scans) were retrieved from two AO-OCT volumes recorded in a young healthy volunteer with a field of view of approximately $4^{\circ} \times 4^{\circ}$ (corresponding to ~$1.4 \times 1.4 mm$ on the retina). Different imaging locations and focus settings were considered. One data set was recorded in the fovea with the focus of the imaging beam set to the posterior retina and one data set close to the optic disc with the focus shifted to the anterior retina. An algorithm including dispersion compensation and geometrical corrections was employed \citep{Shirazi:20} for the reconstruction.
The reference images, (a) in Figure \ref{aooct1} and \ref{aooct2}, were obtained by manually optimizing over the parameters that define the compensation of dispersion, rotation and curvature, respectively. The examples in Figure \ref{aooct1} and \ref{aooct2} compare the chosen reference to three sub-optimal reconstructions, where (b) had a bad choice for the rotation correction parameter, (c) for the curvature correction parameter and (d) for the dispersion compensation parameter. 

\subsubsection*{FR-IQA mismatches}
Good dispersion compensation should provide images with a depth resolution that is optimized for the system at hand. In the ophthalmic application of AO-OCT, this high axial resolution allows for the visualization and identification of the different retinal layers \citep{Shirazi:20,Brunner:21,Wojtkowski:04}, different retinal layers and structures, such as blood vessels. In the first example Figure \ref{aooct1} the clear separation of the different layers in the posterior retina, such as the photoreceptor bands and the retinal pigment epithelium, is crucial for clinicians/researchers who investigate the structure and function of the healthy and the diseased human retina \citep{10.1167/iovs.14-14907}. Therefore, in this example, the reconstruction shown in Figure 2 (d) should have been rated the lowest as the axial resolution is the lowest because of faulty dispersion compensation, which cannot be fixed by further post-processing. 
In the second example Figure \ref{aooct2}, a cross-sectional view of three retinal vessels which are embedded in the nerve fibre layer is given. Changes in the thickness of vessel walls are an important early biomarker for retinal diseases such as diabetic retinopathy, cf.~\citep{BAKKER2022369}. Again, the reconstruction with faulty dispersion compensation shown in Figure \ref{aooct2}(d) should have been rated the lowest. The loss in axial resolution worsens the visualization of the vessel walls and would lead to inaccurate measurements of the vessel wall thickness.

In the current version of the algorithm the parameters which determine the amount of dispersion, rotation and curvature compensation applied by the reconstruction algorithm have to be set manually. Therefore, automated evaluation would be very helpful to fasten the process. PSNR and SSIM are not suitable to assess the problem correctly, as they penalize the spatial deviations in (b) and (c) strongly. The geometric deviations are not beneficial, but these errors could be corrected in an additional post-processing step unlike the deteriorated axial resolution due to the wrong dispersion compensation parameter (d). LPIPS is able to ignore these small spatial deviations. 

\subsection{Digital pathology}
Digital pathology integrates the acquisition, management and interpretation of pathology information generated from digitalized tissue stainings present on glass slides \citep*{digp}. The process starts with high-throughput scanning of glass slides on dedicated slide scanners. The obtained images can be further used for diagnostics, web-based consultations with other expert pathologists in tumour boards, quantitative analysis, and secure archival of pathology data as well as for the development of machine learning tools for tumour classification. However, the application of the listed operations is only valuable upon high image quality and a cost-effective scanning process. Image quality is highly dependent on scanner type and scan settings.
Furthermore, image size is considerably large, with one image usually comprising 1-2 GB, such that data storage and data access represent additional challenges for digital pathology. In this light, optimizing the digitization process through the establishment of quantitative criteria for image quality standards would greatly contribute to efficient workflows and manageable image usage. 

\subsubsection{Visualization Problem}
Currently, the slide scanner operators personally set the parameters on the machine to optimize the scanning procedure. At first, the scanning device performs a low-resolution overview scan in an automated manner. The operator is required then to select a dedicated field of imaging for high-resolution imaging with a 40X objective, followed by the application of focus points to the selected scan area. The distribution of these focus points can either be performed in an automated manner by the scanner software (up to 9 focus points), or manually by the operator (if more than 9 focus points are desired). The choice of how many focus points are used is usually based on the operator’s personal preference, where setting focus points manually is highly time-consuming. Furthermore, each focus point increases the required scanning time for slide digitization, thus adding to the time investment of pathologists at the machine while at the same time limiting the number of slides that can be scanned within one day. 
However, image quality correlates with the ability of the pathologists to provide an accurate diagnosis to their patients. The diagnosis can have life-impacting consequences, such as the choice of the best treatment options for tumour patients based on tumour subtype classification by the histopathological evaluation of tissue sections. Thus, automated IQA for different choices of focus points would be helpful to design a standardized, cost and time effective workflow while providing medical experts with images of reasonable quality for accurate diagnostics.

\subsubsection{Data}
\begin{figure}
\centering
\subfigure[Reference]{\includegraphics[width=0.3\textwidth]{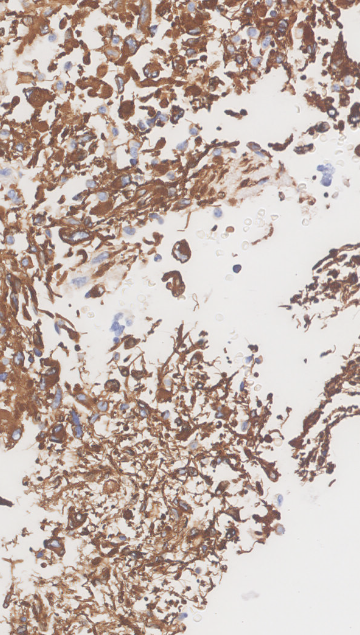}}  
\subfigure[(18.6, 0.80, 0.10)]{\includegraphics[width=0.3\textwidth]{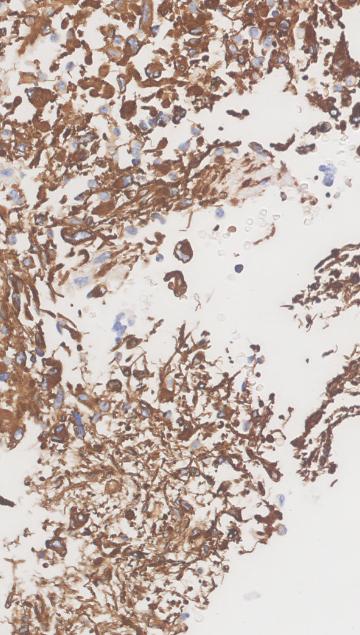}} 
\subfigure[(23.6, 0.91, 0.33)]{\includegraphics[width=0.3\textwidth]{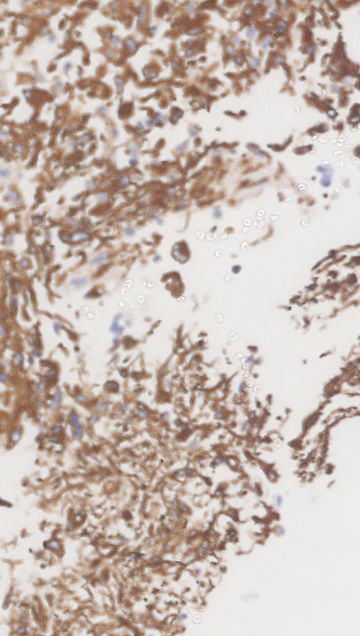}} \\
\subfigure[Reference]{\includegraphics[width=0.3\textwidth]{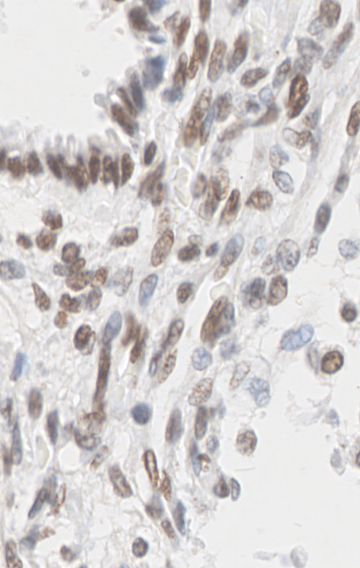}  }
\subfigure[(22.9, 0.81, 0.11)]{\includegraphics[width=0.3\textwidth]{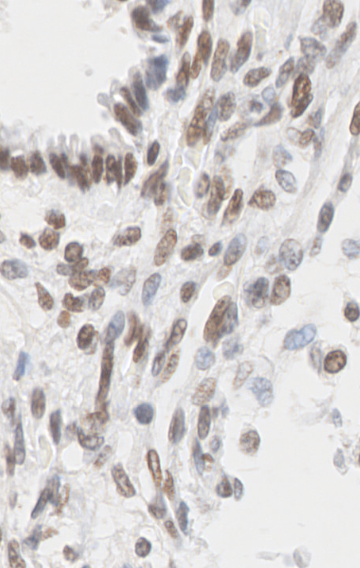}} 
\subfigure[(27.0, 0.87, 0.22)]{\includegraphics[width=0.3\textwidth]{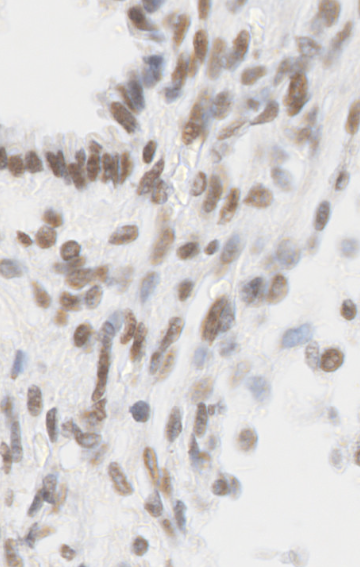}} 
\caption{Data acquired with a slide scanner and 9 (a)(d), 3 (b)(e) and 1 (c)(f) focus points. The image with $9$ focus points serves as a reference here. PSNR and SSIM misjudge the tiny spatial misalignment and therefore favor the blurry images with 1 focus point. LPIPS is able to ignore these spatial misalignments.}
\label{digpath}
\end{figure}

The data presented in this study have been acquired from digital images of immunohistochemistry stainings that were performed on archival tissue obtained from the neurobiobank of the Division of Neuropathology and Neurochemistry at the Medical University of Vienna. Stainings have been performed according to standard procedures \citep{digpathnew1, digpathnew2}. Figure \ref{digpath} (a)-(c) shows a tumour biopsy of a gliosarcoma patient stained for the astrocyte marker GFAP (brown signal, cytoplasmic localization) and counterstained with Hematoxylin (blue signal, nuclear localization). Figure 3 (d)-(e) shows fetal cerebellar tissue stained for the epigenetic mark H3K27me3 (brown signal, nuclear localization) and counterstained with Hematoxylin (blue signal, nuclear localization). 
The stained sections have been digitalized on a NanoZoomer 2.0-HT digital slide scanner C9600 (Hamamatsu Photonics, Hamamatsu, Japan). The corresponding software NPD.Viewer2 was used to export the scanned images to tiff files.
Here, we performed individual scans of a selected imaging area with different numbers of focus points. We chose either 1, 3 or 9 focus points while not changing the spatial settings for the selected field of interest. The image with 9 focus points, allowing the highest resolution, serves as the reference image.

\subsubsection*{FR-IQA mismatches}
Although the spatial settings for the selected scan area of interest were not changed during the experiment, the physical performance of the scanner showed slight spatial deviations of the selected area between individual scans and thus does not allow for high spatial accuracy during re-scanning processes. PSNR and SSIM fail to correctly assess the images in Figure \ref{digpath} as they are very sensitive to that kind of spatial misalignment. Whereas the scan with 3 focus points corresponds much better to the higher-quality reference as the blurred scan with 1 focus point, see (b) and (e) versus (c) and (f), both measures incorrectly judge the blurred scan as the better one. This wrong judgement due to a tiny spatial change is very problematic in the respective framework as it is impossible to guarantee completely exact spatial alignment, even if no other settings had been changed during the scanning process. LPIPS, not being so sensitive to small spatial deviations, is able to correctly judge the rank of quality here.

\subsection{Photoacoustic Data}\label{pa}
Photoacoustic imaging (PAI) is an emerging medical imaging modality with important clinical applications such as inflammatory bowel disease and cardiovascular diseases~\citep{Janek1,Janek2}. PAI combines ultrasound with optical imaging to break through the optical diffusion limit, enabling imaging at depths beyond the reach of conventional optical methods. The inherent optical contrast offers valuable insights into tissue composition and function, enabling enhanced visualization of anatomical structures and pathological abnormalities in vascularization~\citep{Janek3} and blood oxygenation~\citep{Janek4}.

\subsubsection{Photoacoustic Inverse Problems}
The inverse problems of PAI pertain to the task to accurately visualize molecular distributions and determine functional tissue information from acquired PA time series signals \citep{Janek5}. It can be broadly divided into two main components: (1) the reconstruction of images from time series measurements (acoustic inverse problem) and (2) the correction for the non-linear light fluence (optical inverse problem). Progress has been made towards both inverse problems, but solutions typically involve forward simulations that rely on assumptions that may not hold for a given clinical application \citep{Janek6}. This makes the reconstruction of images from photoacoustic measurements challenging, especially for medical applications, and an active field of research. To ensure a suitable quality of reconstructed PA images, objective measures would be needed.

In the field of PAI it is extremely difficult to compare new methods to the state of the art, as many algorithms are not available open source and thus have to be reimplemented from scratch. Furthermore, as no standards exist yet, the field uses standard IQA measures, such as PSNR and SSIM, over measures tailored for PA reconstruction problems \citep{ASSI2023100539}. Standardised, objective measures are therefore highly needed, especially for clinical applications to ensure high visual quality of the data used for diagnosis and prognosis. 

\begin{figure}
\centering
\subfigure[Reference 1]{ \includegraphics[width=0.2\textwidth] {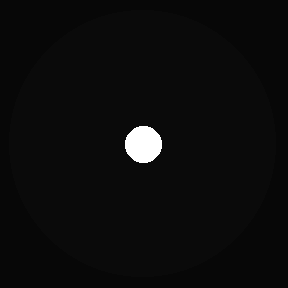}} \ 
\subfigure[(25.1, 0.49, 0.18)]{
  \includegraphics[width=0.2\textwidth]{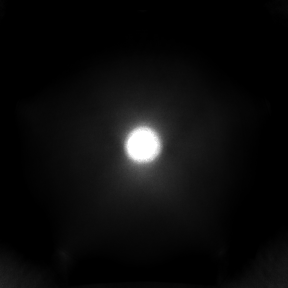}} \ 
\subfigure[(21.3, 0.81, 0.27)]{
  \includegraphics[width=0.2\textwidth]{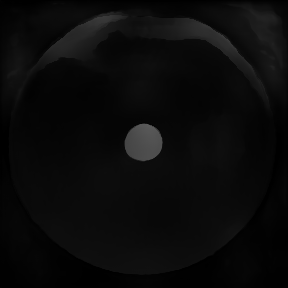}} \ 
\subfigure[(22.5, 0.97, 0.07)]{
  \includegraphics[width=0.2\textwidth]{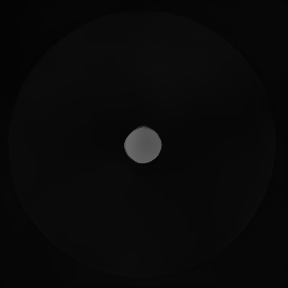}} \\
\subfigure[Reference 2]{ \includegraphics[width=0.2\textwidth]{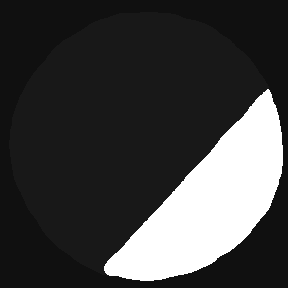}} \ 
\subfigure[(13.1, 0.40, 0.49)]{
  \includegraphics[width=0.2\textwidth]{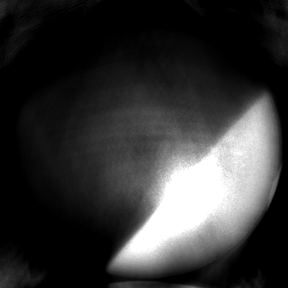}} \ 
\subfigure[(12.5, 0.67, 0.47)]{
  \includegraphics[width=0.2\textwidth]{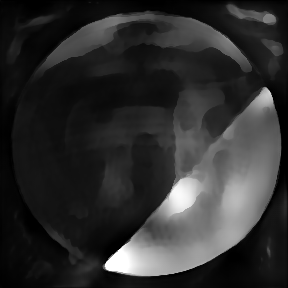}} \ 
\subfigure[(12.5, 0.82, 0.23)]{
  \includegraphics[width=0.2\textwidth]{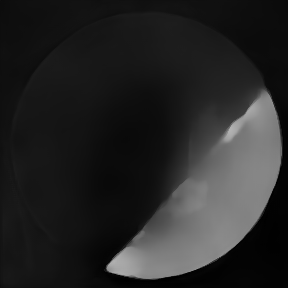}}
  \caption{Two examples of photoacoustic images, ground truth images (a)(e) and comparisons of the reconstructions with \textit{Alg1} (b)(f), \textit{Alg2} (c)(g) and \textit{Alg3} (d)(h). None of the measures was able to assess the relevant quality properties for both examples correctly.}
    \label{photoiqa}
\end{figure}

\subsubsection{Data}
As examples for PA image reconstructions, we show in Figure \ref{photoiqa} reconstructed images containing estimated distributions of the optical absorption coefficient from cross-sectional photoacoustic images of piecewise-constant test objects (phantoms), cf.~\citep{Janek7}. These images were the result of a two-stage process to solve the PA inverse problems. PA data was acquired with a commercial photoacoustic imaging system (MSOT InVision 256-TF, iThera Medical GmbH, Munich, Germany) and processed with three different algorithms (referred to as \textit{Alg1, Alg2, Alg3}). For the visualization and assessment, the outputs of the algorithms were clipped with the reference image's maximum. 

\textit{Alg1} corrects a reconstructed PA image by using the light fluence obtained from simulations based on the reference measurements. \textit{Alg2} and \textit{Alg3} are deep-learning algorithms that are trained to directly estimate the absorption coefficient from the reconstructed image. \textit{Alg2} was trained with simulated data and \textit{Alg3} was trained with experimental data. The reference absorption coefficients are obtained by using a double-integrating sphere \citep{Janek8} setup as a complementary measurement system, which only yields point estimates for homogeneous material samples. Because of the piecewise-constant nature of the used phantoms, one can fabricate an additional batch of the material used for the test object, measure it, and relate the calculated properties to the test object. For any complicated objects or in vivo images, this process would be unfeasible. 

\subsubsection*{FR-IQA mismatches}
Photoacoustic images are typically sparse, which complicates the application of image-based quality measures.
In the original paper \citep{Janek7}, the qualitative assessment was conducted manually. In that assessment, \textit{Alg3} should perform best, \textit{Alg1} second best, and \textit{Alg2} worst across the test images. As this kind of manual assessment is not feasible for larger data sets, and may naturally also introduce inconsistencies as well as biases, automation is needed. However, when PSNR and SSIM are applied to the images these intended results cannot be reproduced. SSIM wrongly assesses that \textit{Alg1} performs worst throughout all both images (Figure \ref{photoiqa}), LPIPS for the second example. PSNR wrongly judges \textit{Alg1} as the best rather than \textit{Alg3} and does not give much lower results for \textit{Alg2} than \textit{Alg3}, despite the introduction of artefacts that lead to significant degradation of image quality. 

Depending on the goal of the data analysis, desired quality properties may differ here. Artifacts might introduce structures into the image that could be mistaken for regions of interest. On the other hand, an inaccurate estimation of the absolute absorption coefficient might lead to errors in the estimation of functional tissue parameters, such as blood oxygenation, that are important to assess the health status of a patient. 
Therefore, targeted IQA measures that are indicative of success given a desired use case are required for the objective assessment of algorithms for quantitative PAI and would allow fast advances in the still evolving field. 

\section{Discussion}\label{discuss}
The results presented in this paper give a collection of examples in diverse medical imaging tasks illustrating the most common pitfalls of the standard FR-IQA measures. Some of the problems associated with these measures have been discussed in isolated studies before. This paper aims to present them in a summarized and structured manner, highlighting the most relevant aspects in the context of medical imaging. 

We showed examples of failure when using the standard FR-IQA measures PSNR and SSIM, as well as the more recently introduced LPIPS (based on AlexNet), which provides some beneficial properties for problems occurring in medical imaging (in particular, invariance to small spatial deviations), but still does not show sufficient performance in many tasks. The occurring problems for the employed measures across the medical imaging modalities are manifold and include

\begin{itemize}
\item penalization of task-irrelevant perceptual information (Figure \ref{CT2}, Figure \ref{xray1}, Figure \ref{xray2} and Figure \ref{photoiqa}),
\item misjudgment of broad artifacts leading to information loss (Figure~\ref{fig:CBCT} and Figure \ref{mri}),
\item inability to detect local errors and structural details (Figure \ref{CT1}), 
\item and for PSNR/SSIM, undesired sensitivity to small spatial changes and geometric deviations (\ref{aooct1}, \ref{aooct2} and Figure \ref{digpath}).
\end{itemize}

The evidence in this paper suggests that there is a need for a comprehensive discussion on standard FR-IQA measures and their applicability in medical imaging tasks. 
Based on the known struggles of FR-IQA measures and the fact that many non-trivial choices have to be made when applying such measures in medical imaging (this includes the choice of image visualization strategy, which strongly influences the result, e.g.~cropping and standardization, as well as implementation details), FR-IQA measures should be acknowledged to not being a straightforward evaluation choice for medical imaging. Moreover, comprehensive reporting about the made choices is needed to ensure reproducibility. 

NR-IQA measures are by design tailored to specific problems and therefore can be applied and interpreted in a more direct way. Unfortunately, such measures or biomarkers are often only known within groups of experts and not always easily accessible for non-experts working outside of the medical field. Recently, research in this direction has evolved, see e.g.~the work on NR-IQA measures for different MRI modalities including the contribution of public data with ratings \citep{MRNR}. Other attempts of clinically tailored NR-IQA measures for MRI include \citep{mriscantime, mrinriqaagain, LEI2022102344}, and for CT data \citep{CTNR, VERDUN2015823, CTmultivendor}. 

Nonetheless, there are medical settings in which FR-IQA measures can be a meaningful addition to NR-IQA measures, e.g.~for quantification purposes in reconstruction algorithms or specific perceptual properties that are needed for the subsequent usage. In order to employ perceptual FR evaluation, more extensive research is highly needed that investigates diverse medical imaging modalities and characterizes targets that can be evaluated well. Such extensive studies should be published independently of final applications in research papers introducing novel methods. In order to enable proper evaluation of the suitability of FR-IQA measures for clinical tasks, medical image data has to be shared, together with expert ratings, such as for NR-IQA the recently introduced CT IQA challenge data set \cite{LEE2025103343}. 

On a related note, a very recent publication \citep{metricsreloaded} provides a comprehensive framework for the choice of imaging analysis metrics for segmentation, detection and classification. Such contributions are very important to bridge the gap between application and methodology through transdisciplinary research, and would also be highly beneficial for IQA measures.  

\subsection{Advancing task-informed medical FR-IQA}
From these thoughts, we can derive an expandable list of research directions that help to advance sufficient evaluation with FR-image quality metrics for medical images:

\begin{enumerate}
\item Studies on existing FR-IQA measures applied to medical data that identify which  quality measures work well, and in which specific settings. The used data and distortions should come from a realistic clinical context, i.e.~degradations should be obtained from medical scanners directly or modeled in a realistic manner. The results may include the realization that some standard measures are working well in some specific tasks; these tasks have to be identified and comprehensively discussed. 
\item Published medical image data with manual expert ratings for specific tasks. Relevant acquisition information should be added, including which kind of visualization and standardization strategies of the images were used in order to obtain the ratings. Did the experts use the same screens? Were they allowed to change the luminance/contrast during the process? 
\item Development of novel task-dependent FR-IQA measures that are tailored and extensively tested for particular medical imaging tasks. 
\item Identification of perceptual quality properties that are needed for specific medical imaging tasks, in connection with perceptual metrics.
\item Development of standard frameworks/platforms to annotate and test FR-IQA measures in a medical setting. 
\end{enumerate}

It is not possible to have one quality measure that serves it all. The usage of task- and image-dependent well-tested measures, e.g.~identified through (1.)~and (3.), is indispensable for sufficient evaluation schemes. However, comparability of results is also of major importance when novel methodologies are introduced. Therefore, in addition, more general measures are needed for broader evaluation of novel methods. 

A way to approach this could be the creation of a carefully curated set of basis measures that correspond to different perceptual properties, e.g.~luminance, contrast, structure and texture. This idea is related to the SSIM framework, which relies on three properties, but eventually combines these to one number. Identified needed perceptual quality metrics (4.) then give subsets for certain tasks. 
Such a basis set can be used for imaging problems arising from several acquisition modalities and may also provide deeper insights into bigger data sets by identifying which kind of quality was preserved well and which has not. 

Most importantly, measures that have not been assessed for medical imaging at all should not be used for such tasks. 
In order to allow assessment of measures for medical imaging tasks, available data (2.) is crucial:  
without shared data, it is not possible to conduct evaluations. 
A standardized framework (5.) would support the implementation of annotation studies. 

Lastly, it should be acknowledged that using perceptual quality measures on acquired medical data is a two-stage problem. Firstly, the question of visualization has to be tackled and secondly, the identification of a suitable perceptual measure. Both of the stages are not trivial for medical images.

\subsection{Guidelines for the choice, description and application of FR-IQA measures}
A set of IQA measures should be reported that includes well tested task-specific FR- and NR-IQA measures as well as visual quality assessment by experts. 
The discussion of suitability of the employed IQA measures should be a comprehensive and important part of a method paper. Whenever possible, qualitative evaluation by experts should be included. 
Moreover, it should not be encouraged to use an IQA measure as a loss function and at the same time as an evaluation metric of a method, as this introduces bias in the optimization/evaluation steps. If not avoidable, the employed measure should be treated and reported as part of the validation but not of the testing phase.

\subsubsection{Guidelines - Application of FR-IQA measures}
In this section, we will suggest guidelines for the usage of FR-IQA measures for the output's evaluation of image processing algorithms with reference data. 
\begin{enumerate}
\item[0.] \textbf{Suitability of IQA} \\
Check if IQA is the correct way to evaluate the image, e.g.~if the image is the final outcome of a process to be visualized or is it just an intermediate result. Verify that the value range corresponds to an image that can be visualized. 
\item \textbf{Locality} \\
Discuss if the measure should be employed on the entire images or if there are specific regions that would ask for distinctive evaluation. Identify a reasonable region of interest and cropping, being aware that huge black background areas influence the outcome.
\item \textbf{Choice of measure set} \\
Verify if the IQA measure is properly tested for the intended kinds of images and tasks. If not, check what specific characteristics of the measure are needed for the evaluation task and if there are other more suitable measures that can be employed. Note if the measure was part of the method optimization. If yes, report it separately as validation measure and add more measures for testing. If possible, add NR-IQA measures that are suitable for the problem and data. If not, specify why.    
\item \textbf{Discuss and argue your choice} \\
Identify properties for your task that shall be tackled by the quality measure, including structure, texture, contrast/brightness, shift (in)variance, color scheme, etc. 
\item \textbf{State exact implementation of the measure} \\ If a measure provides a framework with parameters and/or different implementations (e.g.~SSIM, see introduction), report which parameters and implementations have been used. State on what kind of data type the measure operates on (e.g.~uint, float, etc) and how that fits your data.
\item \textbf{Data standardization} \\ Report conducted data standardization, e.g. how have the images been visualized and what kind of standardization was necessary to visualize the images in a meaningful way. Check if the visualization corresponds to the evaluated FR-IQA quantification. If not, and a perceptual measure was employed, specify why. 
\item \textbf{Share your code} \\ 
Make your method as well as your evaluation schemes publicly available to ensure reproducibility. 

\end{enumerate}

\section{Limitations of the study}
In this paper, we have gathered examples of different medical imaging modalities and identified tasks that commonly use or would benefit from using FR-IQA measures. The choice of examples was made based on relevance, but also guided by the available fields of expertise of the people who span the present collaborative network. This collection is by no means complete and could be widely extended. 

We included the two most commonly used FR-IQA measures for the evaluation of computational medical imaging tasks (PSNR and SSIM) and focused on exemplifying their failures in such settings. There are many tasks in which these measures perform very well, especially for natural images, but evidence is given here that in various medical imaging tasks more extensive testing regarding suitability is necessary. The novel IQA measure LPIPS was also included because there is an ongoing trend to suggest this measure for medical imaging tasks; the message of this paper is to apply caution and conduct further research regarding its broad applicability. The list of existing FR-IQA measures is long, and it is out of the scope of this paper to provide an extensive evaluation scheme of quality measures. Instead, we want to provide examples of failure, highlighting the need for further research regarding the applicability of IQA measures.

\section{Summary}
This study shall serve as a first structured collection of pitfalls in medical imaging tasks when employing the most commonly used FR-IQA measures for evaluation, namely PSNR and SSIM, as well as the more recently introduced deep learning based measure LPIPS. The collection includes examples analyzed by several experts from real-world medical imaging problems with CT, MRI, X-Ray, photoacoustic and pathological image data, where failures may ultimately affect clinical tasks conducted on reconstructed images. Moreover, we formulate guidelines for the application of FR-IQA measures in medical settings and provide suggestions for future research directions. We hope that this critical review will encourage more researchers to actively participate in the research field of evaluation methods and their constraints, in particular regarding FR-IQA. Concluding, this opens up fantastic opportunities for new impactful interdisciplinary research directions where several research communities can benefit from. 

\section{Conflict of Interest}
The authors declare that they have no conflict of interest.

\section{Statements \& Declarations}
\subsection{Funding}
The authors wish to acknowledge support from the EU/EFPIA Innovative Medicines Initiative 2 Joint Undertaking - DRAGON (101005122) (A.Br., I.S., M.R., S.D. C.B.S., AIX-COVNET); the Austrian Science Fund (FWF) through project P33217 (A.Br., C.K.) and through SFB 10.55776/F68,'Tomography Across The Scales', project F6807-N36 (E.B.) and project V1041 (N.A.); the OEAW/L’oreal Austria through the fellowship ’FOR WOMEN IN SCIENCE’ (A.Br.); the U.S. National Science Foundation through grant DMS-2208294 (M.S.L);  the Accelerate Programme for Scientific Discovery and EPRSC grant EP/W004445/1 (A.Bi.); the German Research Foundation through grant GR 5824/1 (J.G.); the National Institute for Health and Care Research (NIHR) Cambridge Biomedical Research Centre (BRC-1215-20014) (I.S.); the EPSRC Cambridge Mathematics of Information in Healthcare Hub EP/T017961/1 (M.R., C.B.S.); and the Trinity Challenge BloodCounts! project (M.R., C.B.S.). 

C.B.S additionally acknowledges support from the Philip Leverhulme Prize, the Royal Society Wolfson Fellowship, the EPSRC advanced career fellowship EP/V029428/1, the EPSRC programme grant EP/V026259/1, and the EPSRC grants EP/S026045/1 and EP/T003553/1, EP/N014588/1, the Wellcome Innovator Awards 215733/Z/19/Z and 221633/Z/20/Z, the European Union Horizon 2020 research and innovation programme under the Marie Skodowska-Curie grant agreement No. 777826 NoMADS, the Cantab Capital Institute for the Mathematics of Information and the Alan Turing Institute. This research was supported by the NIHR Cambridge Biomedical Research Centre (NIHR203312). 
Please note that the content of this publication reflects the authors’ views and that neither NIHR, the Department of Health and Social Care, IMI, the European Union, EFPIA, nor the DRAGON consortium are responsible for any use that may be made of the information contained therein.

\subsection{Competing Interests}
The authors have no relevant financial or non-financial interests to disclose.

\subsection{Ethics approval}
The Brent Research Ethics Committee provided ethical approval for our retrospective X-ray study (IRAS ID: 282705, REC No.: 20/HRA/2504, R\&D No.: A095585). Informed consent was not required as data was pseudonymized.

\section{Acknowledgment}

We also want to acknowledge and thank the members of the AIX-COVNET collaboration: \\
Michael Roberts$^{1,2}$, Sören Dittmer$^{1}$, Ian Selby$^{4,5}$, Anna Breger$^{1,6}$, Matthew Thorpe$^{7}$, Julian Gilbey$^{1}$, Jonathan R. Weir-McCall$^{4,5,8}$, Judith Babar$^{4,5}$, Effrossyni Gkrania-Klotsas$^{2,4}$, Jacobus Preller$^{2}$, Lorena Escudero Sánchez$^{5,9}$, Anna Korhonen$^{10}$ Emily Jefferson$^{11}$, Georg Langs$^{12}$, Helmut Prosch$^{12}$, Guang Yang$^{13}$, Xiaodan Xing$^{13}$, Yang Nan$^{13}$, Ming Li$^{13}$, Jan Stanczuk$^{1}$, Jing Tang$^{14}$, Tolou Shadbahr$^{14}$, Philip Teare$^{15}$, Mishal Patel$^{15,16}$, Marcel Wassink$^{17}$, Markus Holzer$^{17}$, Eduardo González Solares$^{18}$, Nicholas Walton$^{18}$, Pietro Lió$^{19}$, James H. F. Rudd$^{2,4}$, John A.D. Aston$^{15}$, Evis Sala$^{5,9,21,22}$ and Carola-Bibiane Schönlieb$^{1}$. \\

\noindent
\small
$^1$Department of Applied Mathematics and Theoretical Physics, University of Cambridge, Cambridge, UK; $^2$School of Clinical Medicine, University of Cambridge, Cambridge, UK; $^3$ZeTeM, University of Bremen, Bremen, Germany; $^4$Addenbrooke’s Hospital, Cambridge University Hospitals NHS Trust, Cambridge, UK; $^5$Department of Radiology, University of Cambridge, Cambridge, UK; $^6$Center of Medical Physics and Biomedical Engineering, Medical University of Vienna, Austria; $^7$Department of Mathematics, University of Manchester, Manchester, UK; $^8$Royal Papworth Hospital, Cambridge, Royal Papworth Hospital NHS Foundation Trust, Cambridge, UK; $^9$CRUK Cambridge Institute, Cambridge, UK; $^{10}$Language Technology Laboratory, University of Cambridge, Cambridge, UK; $^{11}$Population Health and Genomics, School of Medicine, University of Dundee, Dundee, UK; $^{12}$Department of Biomedical Imaging and Image-guided Therapy, Computational Imaging Research Lab Medical University of Vienna, Vienna, Austria; $^{13}$National Heart and Lung Institute, Imperial College London, London, UK; $^{14}$Research Program in Systems Oncology, Faculty of Medicine, University of Helsinki, Helsinki, Finland; $^{15}$Data Science \& Artificial Intelligence, AstraZeneca, Cambridge, UK; $^{16}$Clinical Pharmacology \& Safety Sciences, AstraZeneca, Cambridge, UK; $^{17}$contextflow GmbH, Vienna, Austria; $^{18}$Institute of Astronomy, University of Cambridge, Cambridge, UK; $^{19}$Department of Computer Science and Technology, University of Cambridge, Cambridge, UK; $^{20}$Department of Pure Mathematics and Mathematical Statistics, University of Cambridge, Cambridge, UK; $^{21}$Advanced Radiodiagnostics Centre, Fondazione Policlinico Universitario Agostino Gemelli, Rome, Italy; $^{22}$Università Cattolica del Sacro Cuore, Rome, Italy.

\newpage

\footnotesize
\bibliography{bib} 
\bibliographystyle{theapa}

\end{document}